\documentclass{aa}

\usepackage{graphicx}
\usepackage{txfonts}
\usepackage{xspace}
\usepackage{tabularx}
\usepackage{supertabular}
\usepackage{longtable}
\usepackage[version=3]{mhchem}
\usepackage[OT2,OT1]{fontenc}
\usepackage{mathtools}
\usepackage{hyperref}
\usepackage{siunitx}
\usepackage{xcolor}
\newcommand\cyr
{
	\renewcommand\rmdefault{wncyr}
	\renewcommand\sfdefault{wncyss}
	\renewcommand\encodingdefault{OT2}
	\normalfont
	\selectfont
}
\DeclareTextFontCommand{\textcyr}{\cyr}

\newcommand{\RNum}[1]{\uppercase\expandafter{\romannumeral #1\relax}}

\newcommand{\perovskite}{CaTiO$_3$}

\newcommand{\WASP}{WASP-121\,b\xspace}
\newcommand{\kms}{km~s$^{-1}$\xspace}

\newcommand{\vmag}{10.4\xspace}
\newcommand{\planetmass}{1.157 \pm 0.070\xspace M_{J}\xspace}
\newcommand{\planetradius}{1.753 \pm 0.036\xspace R_{J}\xspace}
\newcommand{\planetteq}{2358 \pm 52\xspace K\xspace}
\newcommand{\orbitalvelocity}{221.1 \xspace}
\newcommand{\systemicvelocity}{38.3 \xspace}
\newcommand{\Teff}{6335\xspace K\xspace}

\newcommand{\water}{H$_2$O \xspace}

\newcommand{\alltestedspecies}{\ion{Na}{I}, \ion{Mg}{I}, \ion{K}{I}, \ion{Ca}{I}, \ion{Ca}{II}, \ion{Sc}{I}, \ion{Ti}{I}, \ion{Ti}{II}, \ion{V}{I}, \ion{Cr}{I}, \ion{Mn}{I}, \ion{Fe}{I}, \ion{Fe}{II}, \ion{Co}{I}, \ion{Ni}{I}, \ion{Sr}{I}, \ion{Y}{I}, \ion{Zr}{I}, \ion{Nb}{I}, \ion{Ba}{I}, \ion{La}{I}, \ion{Ce}{I}, AlO, TiO and VO\xspace}

\newcommand{\alldetectedspecies}{\ion{Ca}{I}, \ion{V}{I}, \ion{Cr}{I}, \ion{Mn}{I}, \ion{Fe}{I}, \ion{Co}{I} and \ion{Ni}{I}\xspace}
\newcommand{\previousspecies}{\ion{H}{I}, \ion{Li}{I}, \ion{Na}{I}, \ion{Mg}{I}, \ion{K}{I}, \ion{Ca}{I}, \ion{Ca}{II}, \ion{Sc}{II}, \ion{V}{I}, \ion{Cr}{I}, \ion{Mn}{I}, \ion{Fe}{I}, \ion{Co}{I} and \ion{Ni}{I} \xspace}

\newcommand{\outlierfilterwidth}{20 pixels\xspace}
\newcommand{\outliercthresh}{$5\sigma$\xspace}

\newcommand{\totalhealed}{$6.32 \times 10^5$ \xspace}
\newcommand{\totalhealedpercentage}{0.166\%\xspace}
\newcommand{\totalmasked}{$9.72 \times 10^6$\xspace}
\newcommand{\totalmaskedpercentage}{2.56\%\xspace}
\newcommand{\telluricthresh}{20 \%}
\newcommand{\velocityrange}{$\pm$ 1000 km~s$^{-1}$\xspace}
\newcommand{\velocitystep}{1 km~s$^{-1}$\xspace}

\newcommand{\exptime}{300 s\xspace}

\begin{document}

   \title{The Mantis Network \RNum{4}:}

   \subtitle{A titanium cold-trap on the ultra-hot Jupiter \WASP.}

   \author{H.J. Hoeijmakers\fnmsep
          \inst{1}\thanks{Corresponding author}
          D. Kitzmann\inst{2}
          \and
          B.M. Morris\inst{3}
          \and
          B. Prinoth\inst{1}
          \and
          N. Borsato\inst{1}
          \and
          B. Thorsbro\inst{1}
          \and
          L. Pino\inst{4}
          \and
          E. K. H. Lee\inst{2}
          \and
          C. Ak{\i}n\inst{2}
          \and
          J.V. Seidel\inst{5}
          \and
          J.L. Birkby\inst{6}
          \and
          R. Allart\inst{7}\thanks{Trottier Postdoctoral Fellow}
          \and
          K. Heng\inst{8,9,10}
          }

   \institute{Lund Observatory, Division of Astrophysics, Department of Physics, Lund University, Box 43, 221 00 Lund, Sweden\\
              \email{jens.hoeijmakers@fysik.lu.se}
         \and
            Center for Space and Habitability, University of Bern, Gesellschaftsstrasse 6, CH-3012 Bern, Switzerland
        \and
        Space Telescope Science Institute, 3700 San Martin Dr, Baltimore, MD 21218, USA
        \and
        INAF – Osservatorio Astrofisicodi Arcetri, Largo Enrico Fermi 5, 50125 Firenze, Italy
        \and
        European Southern Observatory, Alonso de Córdova 3107, Vitacura, Región Metropolitana, Chile
        \and
        Astrophysics, Department of Physics, University of Oxford, Keble Road, Oxford, OX1 3RH, UK
        \and
        D\'epartement de Physique, Institut Trottier de Recherche sur les Exoplan\`etes, Universit\'e de Montr\'eal, Montr\'eal, Qu\'ebec, H3T 1J4, Canada
        \and
        University Observatory, Faculty of Physics, Ludwig Maximilian University, Scheinerstrasse 1, D-81679 Munich, Germany
        \and
        ARTORG Center for Biomedical Engineering Research, Murtenstrasse 50, CH-3008, Bern, Switzerland
        \and
        Astronomy \& Astrophysics Group, Department of Physics, University of Warwick, Coventry CV4 7AL, United Kingdom
        }

   \date{Received September 13, 2022; Accepted October 6, 2023}


  \abstract
   {Using emission lines by metals we investigate the three-dimensional distribution of temperature and chemistry in ultra-hot Jupiters.}
   {Existing observations of \WASP have suggested an under-abundance of titanium and titanium-oxide from its terminator region. In this study, we aim to determine whether this depletion is global by investigating the day-side emission spectrum.}
   {We analyze eight epochs of high-resolution spectra obtained with the ESPRESSO spectrograph, targeting orbital phases when the day-side is in view. We use a cross-correlation method to search for various atoms, TiO and VO and compare to models. We constrain the velocities and phase-function of the emission signal using a Bayesian framework.}
   {We report significant detections of \alldetectedspecies, but not Ti or TiO. Models containing titanium are unable to reproduce the data. The detected signals are consistent with the known orbital and systemic velocities and with peak emission originating from the sub-stellar point.}
   {We find that titanium is depleted from regions of the atmosphere where transmission and emission spectroscopy are sensitive. Supported by recent HST observations of the night-side, we interpret this as evidence for the night-side condensation of titanium, preventing it from being mixed back into the upper layers of the atmosphere elsewhere on the planet. Species with lower condensation temperatures are unaffected, implying sharp chemical transitions exist between ultra-hot Jupiters that have slight differences in temperature or dynamical properties. As TiO can act as a strong source of stratospheric heating, cold-trapping creates a coupling between the thermal structures on the day-side and night-side, and thus condensation chemistry needs to be included in global circulation models. Observed elemental abundances in hot Jupiters will not reliably be representative of bulk abundances unless night-side condensation is robustly accounted for or the planet is hot enough to avoid night-side cold-traps entirely. Secondary eclipse observations by JWST/NIRISS have the potential to confirm an absence of TiO bands at red-optical wavelengths. We also find that the abundance ratios of metal oxides to their atomic metals (e.g. TiO/Ti) depend strongly on the atmospheric C/O ratio, and that planetary rotation may significantly lower the apparent orbital velocity of the emission signal.}

   \keywords{planets and satellites: atmospheres -- planets and satellites: composition -- planets and satellites: gaseous planets -- methods: observational -- techniques: spectroscopic}

   \maketitle
%





\section{Introduction}
\WASP is a highly inflated ($R=\planetradius$, $M=\planetmass$), highly irradiated ($T_{\textrm{eq}}=\planetteq$) exoplanet orbiting a relatively bright ($V=\vmag$) F6V star with a short orbital period of 1.27 days \citep{Delrez2016,Bourrier2020}. Because of its high temperature, it is classified \citep{Evans2017} as a member of the class of ultra-hot Jupiters (UHJ). The atmospheres of UHJs are theorized to be largely atomic, as all but the most strongly bound molecules dissociate \citep{Arcangeli2018,Kitzmann2018,Lothringer2018,Parmentier2018}. As the system is observationally favorable, \WASP has been targeted intensively over the past years with several ground-based and space-based facilities, including optical and infrared spectrographs to study the planet's atmosphere. Initially, it has been observed extensively with the WFC3 and STIS instruments on the Hubble Space Telescope (HST). These observations revealed emission by \water \citep{Evans2017} that is indicative of a thermal inversion, now commonly observed in ultra-hot Jupiters \citep[e.g.][]{Nugroho2020,Pino2020,May2021}. The optical transmission spectrum appears to be complex, with the absorption of metals and possibly TiO or VO leading to significant variability in the transit radius between near-UV, optical and near infra-red wavelengths \citep{Evans2016,Evans2018}. Recent phase-curve observations by \citet{Evans2022} have placed tight constraints on the temperature profiles of the day and night-sides, as well as the metallicity, which appears to be elevated.

Numerous transit observations have also been obtained using ground-based high-resolution spectrographs \citep{Hoeijmakers2020,Merritt2021,Borsa2021}. With spectral resolving powers of $R\sim10^5$, such instruments resolve individual absorption lines of atoms and molecules in the spectra of exoplanets, and measure Doppler shifts down to less than 1 km/s, enabling observations of the effects of atmospheric dynamics \citep{Snellen2010}. As the transmission spectra of UHJs are typically rich in metal absorption lines, optical high-resolution spectrographs have proven to be highly capable of identifying metals in the atmospheres of this type of planet, including \WASP, by use of the cross-correlation technique \citep{Snellen2010} that combines the contributions of many individual lines. Observations taken with ESO's HARPS, UVES and ESPRESSO spectrographs have revealed the following species: \previousspecies \citep{Bourrier2020,Cabot2020,Hoeijmakers2020, Merritt2021, Borsa2021}. Ti and TiO were predicted to cause observable absorption and have been searched for repeatedly, without either of them being detected \citep{Hoeijmakers2020,Merritt2021,Gibson2022}. Recent advances in the application of the high-resolution cross-correlation technique have allowed for the retrieval of elemental abundances and abundance ratios from this type of data \citep{Gibson2022}, also confirming the depletion of titanium compared to other metals at the terminator region. The apparent absence of Ti and TiO are consistent with the HST STIS \& WFC3 transmission spectra, which favor VO bands combined with a depletion of TiO \citep{Evans2018}. It has been hypothesized that the depletion of Ti from the observable part of the atmosphere is caused by condensation of Ti out of the gas phase \citep{Hoeijmakers2020,Evans2022}. A question that arises is where in the atmosphere this condensation of Ti occurs and to what extent it is distributed around the planet. Are Ti and TiO missing from the transmission spectrum because Ti is condensing locally at the terminator region, or because condensation happens at a cold point elsewhere on the planet (e.g. the night-side), where it remains locked up in what is known as a cold-trap \citep{Spiegel2009}. The observation of V and the simultaneous absence of Ti/TiO suggests that this condensation mechanism is ineffective for vanadium, placing the temperature profile in the coldest regions of the atmosphere between the stability curves of \perovskite\ and VO, which are the highest-temperature condensates of Ti and V respectively at solar element abundances \citep{Lodders2002}. The peak day-side temperature of \WASP has most recently been measured to be over 3,000\,K \citep{Evans2022}, well above the condensation temperature of \perovskite. The same observations constrained the night-side temperature to be near or below the condensation curve of \perovskite, supporting the hypothesis that titanium can indeed condense on the night-side.

The objective of this study is to present observational evidence that titanium is missing from the day-side emission spectrum of the planet, implying that it is depleted globally and thus indeed cold-trapped. Section \ref{sec:obs_and_data} describes our ESPRESSO observations and the data reduction and pre-processing. Section \ref{sec:xcor_method} describes the analysis methodology, based on the high-resolution cross-correlation technique \citep{Snellen2010}. Section \ref{sec:results} summarizes and discusses the results and important implications.

\section{Observations and data reduction}\label{sec:obs_and_data}

We observed the WASP-121 system (spectral type F6, V=10.4) during 8 epochs with the ESPRESSO instrument at ESO's 8.2 m VLT (PID 0105.C-0591 \& 0106.C-0737, P.I.\, Hoeijmakers), yielding 8 continuous time-series of the high-resolution spectrum of the system. The observations were carried out from January to April 2021, at various phases of the planet, with four epochs before and four epochs after the secondary eclipse to cover phase ranges in which the planet accelerates in opposing directions. The observations are summarized in Table \ref{tab:obs}, and the resulting raw data can be accessed publicly from ESO's Science Archive Facility. The observations used the standard calibration plan of the instrument, including flat-field and wavelength calibrations. In all cases, the instrument was used in 1-UT, high-resolution mode with \exptime exposures, yielding spectra with a resolving power of $R\sim 140,000$) and a typical SNR of 50. The instrument was set to a readout mode of 2x1 binning with fiber B on sky.

\begin{table}
\centering
\caption{Overview of the observing epochs, where $N_{\mathrm{exp}}$ is the number of exposures taken during each run, $\phi$ is the range of orbital phases of \WASP covered and weight is the relative weight of each night, when used to combine the cross-correlation functions in Section \ref{sec:results}. In particular the fifth night is weighed down because those observations covered orbital phases near quadrature where the change in radial velocity is small, leading to partial self-subtraction of the planetary signal.}
\begin{tabular}{ccccc}
\hline \hline
\# & Date & $N_{\mathrm{exp}}$ & $\phi$ & weight \\
\hline
1 & 2021-01-11 & 41 & 0.272 -- 0.403 & 0.211 \\
2 & 2021-01-23 & 37 & 0.563 -- 0.684 & 0.156 \\
3 & 2021-02-06 & 37 & 0.569 -- 0.694 & 0.165 \\
4 & 2021-03-21 & 43 & 0.225 -- 0.370 & 0.099 \\
5 & 2021-04-04 & 33 & 0.190 -- 0.307 & 0.026 \\
6 & 2021-04-07 & 28 & 0.575 -- 0.667 & 0.059 \\
7 & 2021-04-08 & 37 & 0.326 -- 0.447 & 0.190 \\
8 & 2021-04-16 & 34 & 0.603 -- 0.715 & 0.096 \\
\hline
\end{tabular}
\label{tab:obs}
\end{table}

\begin{figure*}
        \includegraphics[width=\textwidth]{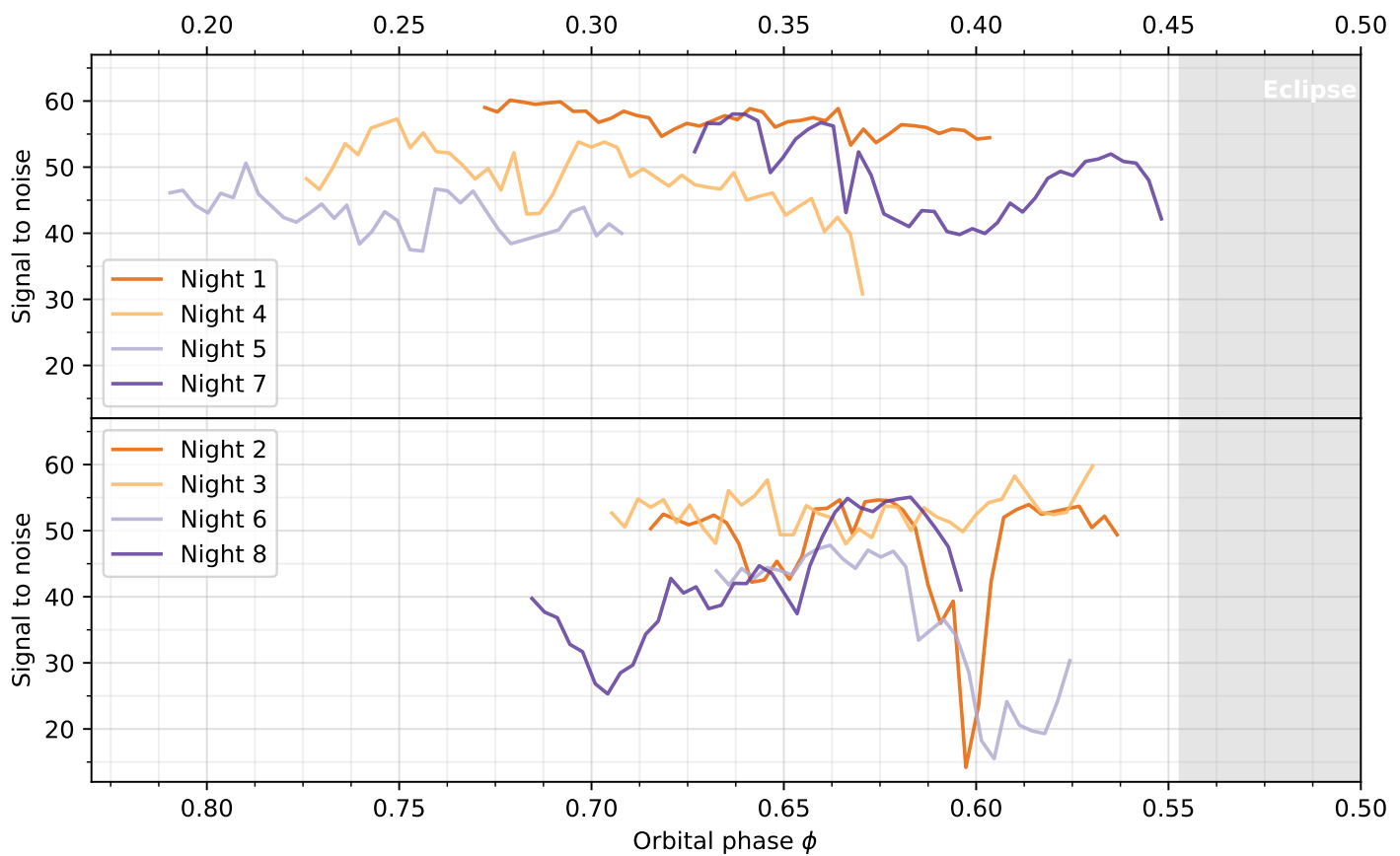}
        \caption{The signal-to-noise ratio and orbital phase of each of the eight time-series. The top-panel shows pre-eclipse data and the bottom panel shows post-eclipse data with the phase axis reversed, such that the secondary eclipse is on the right (highlighted in gray). The signal-to-noise ratio is as measured by the DRS in order 102, covering wavelengths around 550 nm. The time-series are numbered chronologically, corresponding to Table \ref{tab:obs}}
         \label{Fig:SNR}
  \end{figure*}

We reduced the data using the ESPRESSO Data Reduction Software (DRS) version 2.3.3. The DRS produces extracted echelle orders (S2D), as well as stitched, resampled and blaze-corrected spectra (S1D) spanning the full wavelength range of the instrument. The signal-to-noise ratios achieved during each of these observations as estimated by the pipeline are shown in Fig. \ref{Fig:SNR}. Throughout the analysis, we use the extracted spectra from the science fiber (A) without sky subtraction to preserve the blaze function and with it, the true flux recorded by the instrument.

\section{Data processing and cross-correlation analysis}\label{sec:xcor_method}
We proceed to investigate the high-resolution spectrum of \WASP using a cross-correlation-based approach that closely follows the methodology used in earlier work where we analyzed high-resolution time-series spectra of transiting UHJs \citep[e.g.][]{Hoeijmakers2020,Prinoth2022}, and is summarized here with emphasis on aspects that are particular to ESPRESSO.

\subsection{Telluric correction}
We model the telluric transmission spectrum by using \texttt{molecfit} version 1.5.9, applied to each of the S1D spectra \citep{Smette2015} \footnote{\texttt{molecfit} version 1.5.9 is the last version of \texttt{molecfit} that can be installed stand-alone. We modified its GUI-environment to be \texttt{python3}-compatible and removed one duplicate entry in the line-list of water near 580 nm. We are willing to redistribute our version of this package upon reasonable request.}. The fitting of the telluric models to the S1D spectra is carried out in the Earth's rest-frame by reversing the barycentric velocity correction that is applied by the ESPRESSO pipeline by default. These are then Doppler-shifted, re-interpolated and divided out of the S2D spectra, yielding telluric-corrected time-series spectra for each of the echelle orders.

\subsection{Post-processing}\label{sec:postprocess}
The telluric-corrected spectra are aligned with the inertial frame of the star by shifting the wavelength solution back to the barycentric rest-frame and removing the Keplerian component caused by the gravitational pull of the planet, as predicted using the known semi-major radial velocity amplitude $K$ and the orbital ephemeris \citep{Bourrier2020}. After these velocity corrections, the spectra are re-interpolated onto a common wavelength grid. This is the only re-interpolation of the extracted orders carried out during this analysis, and is done to allow for the spectral orders to be treated as 2D variables (wavelength versus time) from here on.

For each order\footnote{Because the fiber is sliced into two halves, each spectral order is recorded twice by the detector. In this analysis we treat each copy as though it was an independent spectral order with its own wavelength and time axes.} we determine the time-average flux and divide it to normalize the order to its average flux; while saving the relative flux differences for later use as weights.
We then determine the time-average spectrum and divide it out to obtain nearly flat residuals, to which we apply color correction and outlier rejection. Color correction is done by fitting a $3^{\rm{rd}}$ order polynomial to each exposure in the time series and dividing it out. This removes broad-band time-variability in the spectrograph's effective throughput, such as variability due to atmospheric dispersion. On these residuals we apply a running median average deviation (MAD) filter with a width of \outlierfilterwidth wavelength columns, to approximate the standard deviation in the order as a function of wavelength. All flux values more than \outliercthresh away from the local median are flagged as bad data. All columns with more than 20\% of pixels flagged as such are instead flagged entirely. All other outliers are replaced by linear interpolates. We then visually inspect all orders and flag remaining bad columns, e.g. due to the presence of telluric emission lines (e.g. sodium) or regions with very low flux at some order edges. We also reject columns where telluric lines absorb more than \telluricthresh\ of the flux. In total, \totalhealed (\totalhealedpercentage) values are replaced by interpolates and \totalmasked (\totalmaskedpercentage) are in columns that are masked entirely. These columns are ignored in the remainder of the analysis.

\subsection{Cross-correlation templates and models}\label{sec:models}

\begin{figure}
        \includegraphics[width=\columnwidth]{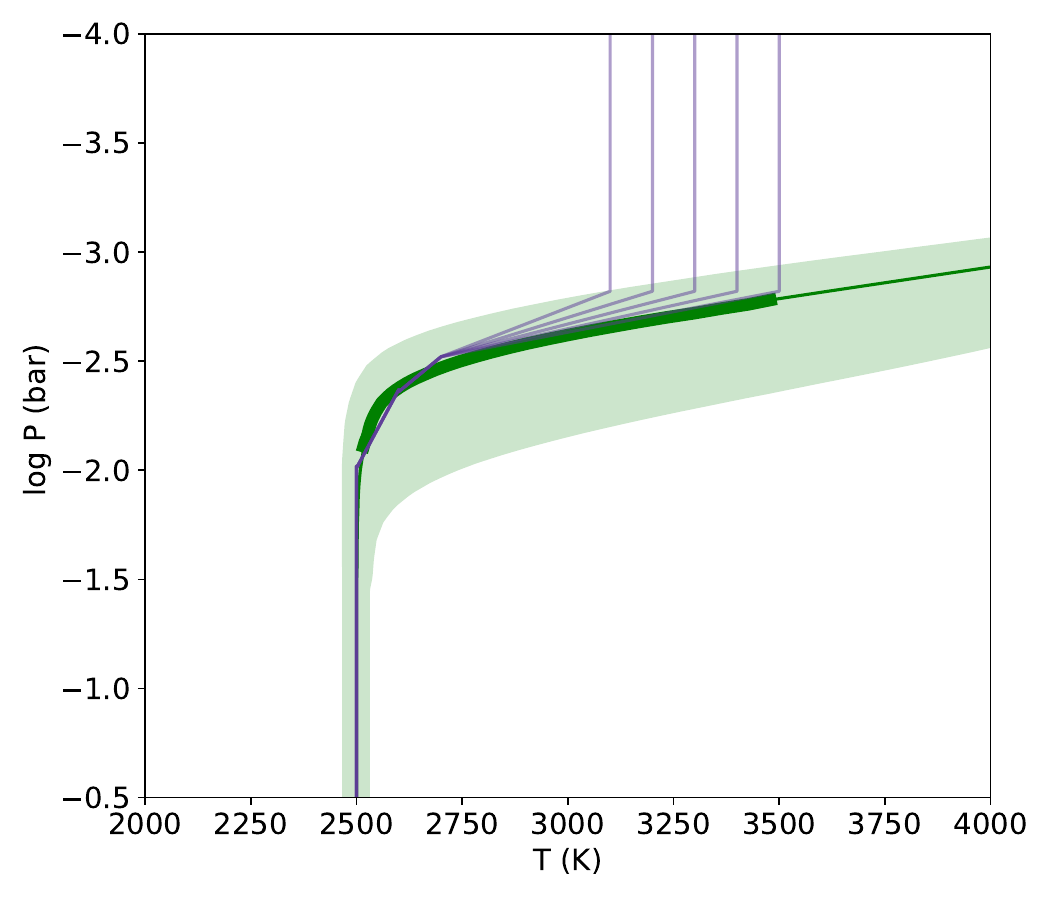}
        \caption{The temperature-pressure profiles used in this work (purple), compared to the best-fit profile by \citet{Evans2022} (green). The thick green line indicates the area where the HST data is sensitive. We chose to explore a variety of isothermal profiles at lower pressures. The T-P profile as plotted in \citet{Evans2022} was digitized using WebPlotDigitizer, \texttt{https://github.com/ankitrohatgi/WebPlotDigitizer}}
         \label{Fig:TP}
  \end{figure}

We created grids of 1D models of the planet emission spectrum for use as cross-correlation templates, as well as to allow for model comparison. The principal free parameter in models of the thermal emission spectrum is the temperature-pressure (T-P) profile. Our choice of T-P profile is based on the recent analysis by \citet{Evans2022}, who find that the emission spectrum is well reproduced with a 1D model with a temperature of 2,500\,K in the deep atmosphere that increases to 3,500\,K at pressures between  10 and 1\,mbar \citet[see Fig. \ref{Fig:TP} as well as Fig. 3 in ][]{Evans2022}. At pressures below 1\,mbar, their best-fit model increases towards 4,000\,K but these pressures are not constrained by the HST data. We therefore construct a class of models that follow the inversion as constrained by \citet{Evans2022}, but that becomes isothermal for $\log P < -2.8$ at a temperature that we vary from 3,100\,K to 3,500\,K towards the top of the atmosphere (see Fig. \ref{Fig:TP}). Later in the analysis, we find that models with peak temperatures between 3,200\,K and 3,300\,K are best capable of explaining our high-resolution data. Our models further assume a metallicity value of either $5 \times$ (0.7 dex) or $10 \times$ solar (1.0 dex) based on the findings by \citet{Evans2022} that the metallicity of \WASP is likely elevated. As in previous work \citep{Hoeijmakers2020}, we assume chemical equilibrium and model the composition of the atmosphere given a value for the metallicity using FastChem 2.0 \citep{Stock2022}. Radiative transfer is done following the procedure in \citep{Gaidos2017}, with line opacity by 151 atoms ions, 8 molecules and continuum opacity caused by Rayleigh scattering, collision-induced absorption and continuum absorption \citep[see ][for details]{Kitzmann2021}. Opacity functions are computed with HELIOS-K 2.0 \citep{Grimm2021} based on line lists adopted from VALD and Kurucz for atoms \citep{Kurucz2017,Pakhomov2019,Ryabchikova2015}, and Exomol for molecules \citep{Barber2014,Li2015,McKemmish2016,Polyansky2018,Bernath2020,Bowesman2021,Syme2021}, including the recently updated TiO line-list \citep{McKemmish2019}. Model spectra are computed in units of spectral flux density (erg s$^{-1}$ cm$^{-2}$ $\mu$m$^{-1}$), and are converted to contrast by dividing by the flux density of the star, modeled as a \Teff blackbody \citep{Polanski2022}, and then multiplying with the square of the radius ratio of the planet and the star.

\begin{figure}
    \includegraphics[width=\columnwidth]{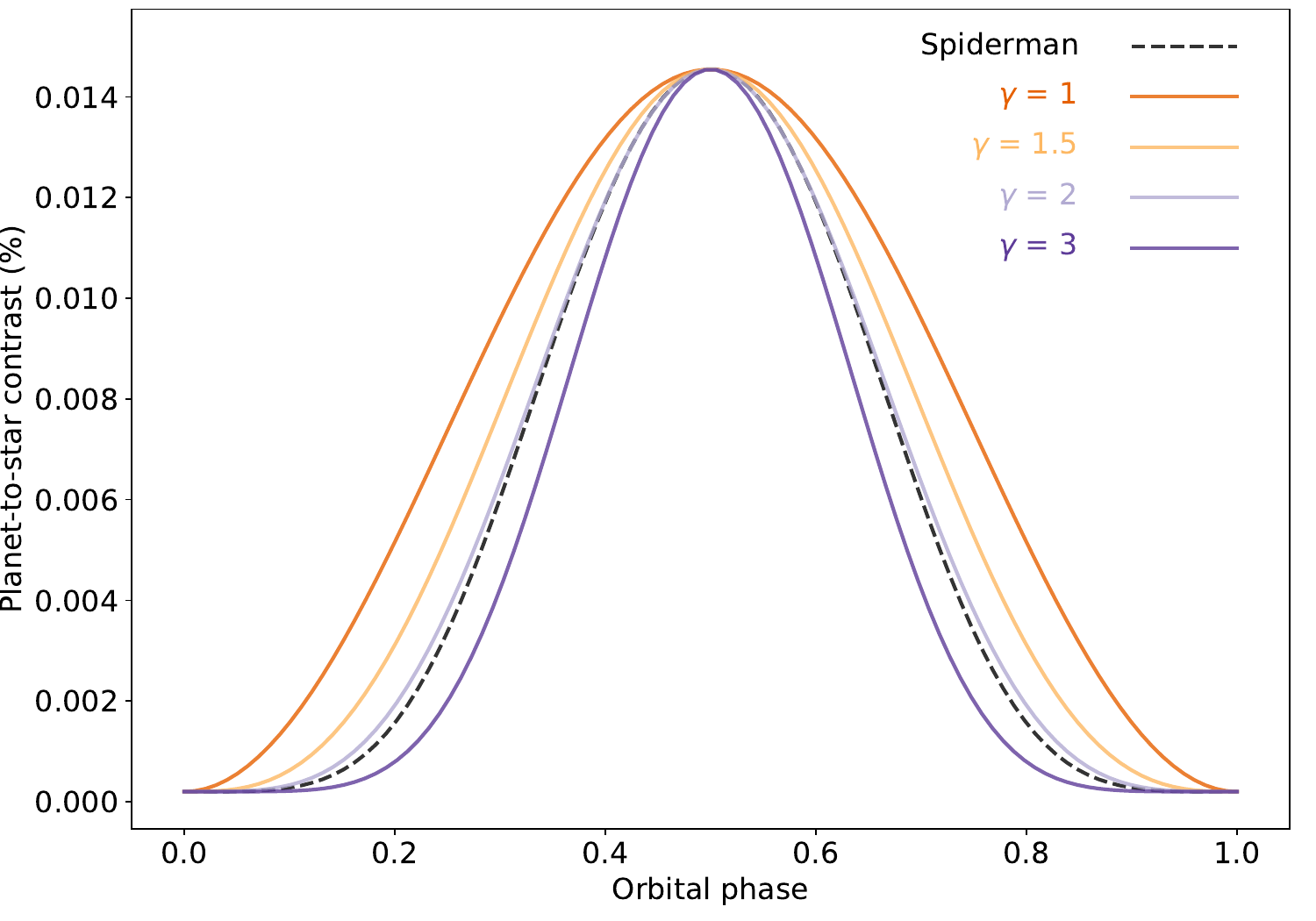}
      \caption{Model of the broad-band phase-curve of \WASP with day-side and night-side temperatures of 3200 K and 1800 K, instantaneous re-radiation integrated over a uniform pass-band from 400 to 800 nm, compared to our parameterized phase-function of equation \ref{eq:phasefunction}.}
         \label{Fig:phasecurves}
  \end{figure}

Models containing all opacity sources are injected into the data by multiplication with a weight that is a function of orbital phase following \citet{Herman2022}, as:

\begin{equation}
    f(\lambda)_{\mathrm{injected}} = f_{\mathrm{obs}}(\lambda) f(\lambda)_{\mathrm{model}} \cdot 0.5 \cdot (1+\cos(2 \pi (\phi-\theta)- \pi)^{\gamma}),
\end{equation}\label{eq:phasefunction}

where $f_{\mathrm{obs}}(\lambda)$ is the data after the masking of bad columns (see above), $f(\lambda)_{\mathrm{injected}}$ is the data with the injected model $f(\lambda)_{\mathrm{model}}$, $\phi$ is the orbital phase, $\theta$ is an offset between the sub-stellar point and the peak of the emission, and $\gamma$ acts to steepen the phase function near secondary eclipse \footnote{For $\gamma=1$, this expression is equivalent to the expression of \citet{Herman2022}, with a contrast parameter $C=1$ and a scaling factor $\alpha = 1$.}. Values of $\gamma = 1$ and $\theta=0$ are adopted for model injection and comparison, as later we will find that the data do not support significant deviations from these values. We also compare this phase-function to the broad-band light-curve of \WASP modeled using \texttt{Spiderman} \citep{Louden2018}, assuming day-side and night-side temperatures of 3200 K and 1800 K, instantaneous re-radiation, the known system parameters \citep{Bourrier2020} and a uniform pass-band from 400 to 800 nm. We find that this phase-curve is well approximated for values of $\gamma \approx 2$ (see Fig. \ref{Fig:phasecurves}, but we note that at present, our high-resolution data does not constrain $\gamma$ very accurately, nor should it be expected that line-emission is described by the same phase-function as broad-band continuum emission.

Cross-correlation templates are constructed by including the opacity of continuum absorbers and only a single species at each time, with a peak temperature of 3,200\,K. A model without any line absorbers is subtracted to obtain a continuum-subtracted template for each species. We perform cross-correlation \citep[equation 1 in ][]{Hoeijmakers2020} using templates for \alltestedspecies, over a velocity range of \velocityrange with steps of \velocitystep on each of the eight time-series (epochs). The resulting cross-correlation time-series are each co-added to the rest-frame of the planet, by assuming a range of possible values for the orbital velocity $K_p$ while fixing the time of transit center \citep{Bourrier2020} and weighing by the flux in the exposures (as determined before in section \ref{sec:postprocess}).

\section{Results and discussion}\label{sec:results}
\begin{figure*}
        \includegraphics[width=\textwidth]{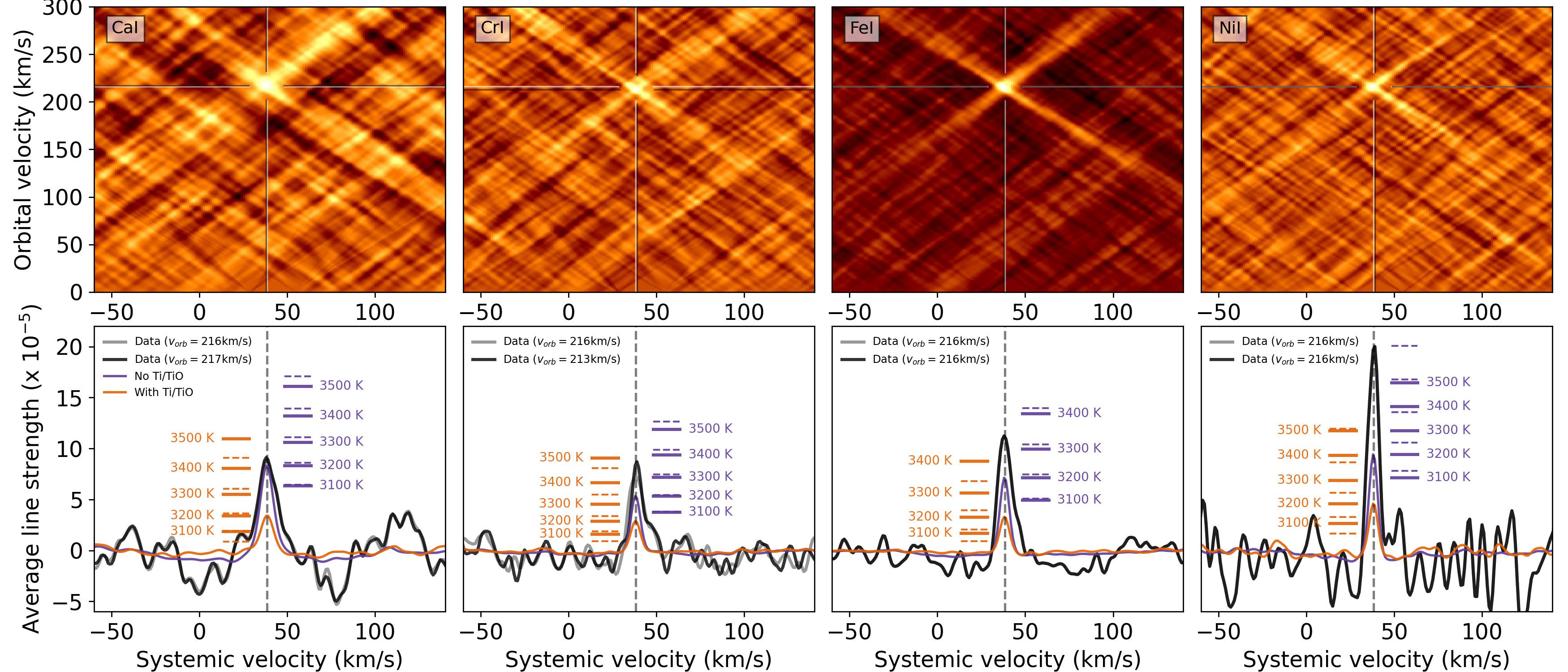}
        \caption{Detections of Ca, Cr, Fe and Ni in the emission spectrum of WASP-121\,b in velocity-velocity space (upper panels) and the extracted one-dimensional CCFs (bottom panels). The one-dimensional CCFs are extracted at an orbital velocity of 216.0 km/s (gray line) slightly below the expected velocity of $\orbitalvelocity$ km/s as derived from the known orbital parameters, or at the orbital velocity at which the peak occurs (black line), to illustrate the effect of the choice of orbital velocity at which to extract. The vertical axis indicates weighted emission line-strength (not relative signal-to-noise). This quantity derives its physical meaning from comparison with injected models. Colored dashed lines indicate the signals caused by injected models with and without Ti/TiO at a metallicity of 5$\times$ solar \citep[equal to 0.7 dex, the metallicity found and used by][]{Evans2022}, with an inverted temperature-pressure profile that is isothermal at 3200 K at pressures of $\log(P) = -2.8$ and below. At altitudes above this pressure, the HST data published by \citep{Evans2022} do not constrain the TP-profile. Horizontal bars indicate the signal strength of models with variations of this peak temperature, and thus the strength of the thermal inversion. The solid bars indicate models with a metallicity of 5$\times$ solar, and the dashed bars indicate 10$\times$ solar.}
         \label{Fig:CCF_results_1}
  \end{figure*}

\begin{figure*}
    \includegraphics[width=\textwidth]{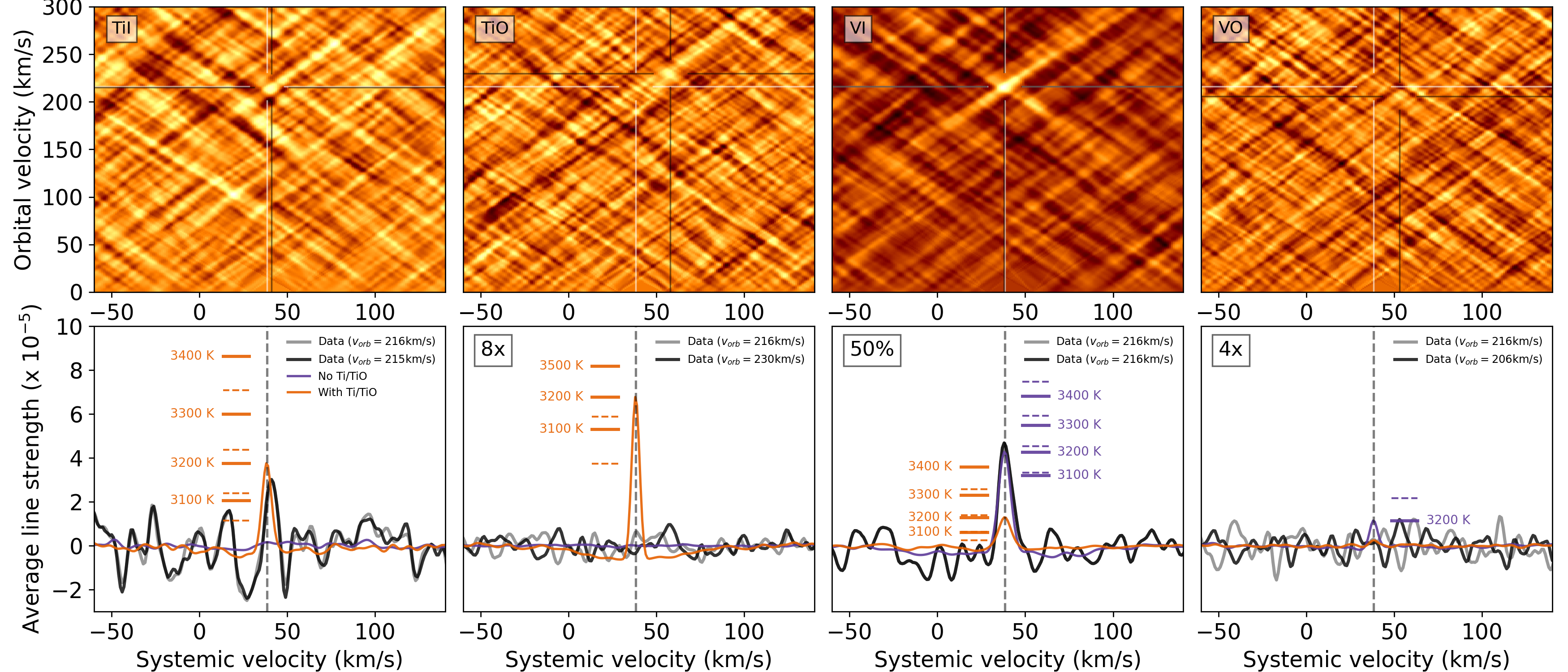}
      \caption{Similar to Fig. \ref{Fig:CCF_results_1}, for Ti, TiO, V and VO. Ti and TiO are not detected but are predicted by models that contain Ti/TiO opacity. VO is not detected nor does the model indicate that the data are sensitive regardless of temperature, unless the metallicity is significantly increased. The horizontal bars of VO for different temperatures overlap so only 3200 K is shown. In-set scaling factors denote the factor by which the y-axis was zoomed, to allow the signals to be plot on the same scale. 50\% means that the vertical axis labels should be read as being a factor of 2 bigger, and vice versa.}
         \label{Fig:CCF_results_2_TiTiO}
  \end{figure*}

\begin{figure*}
    \includegraphics[width=\textwidth]{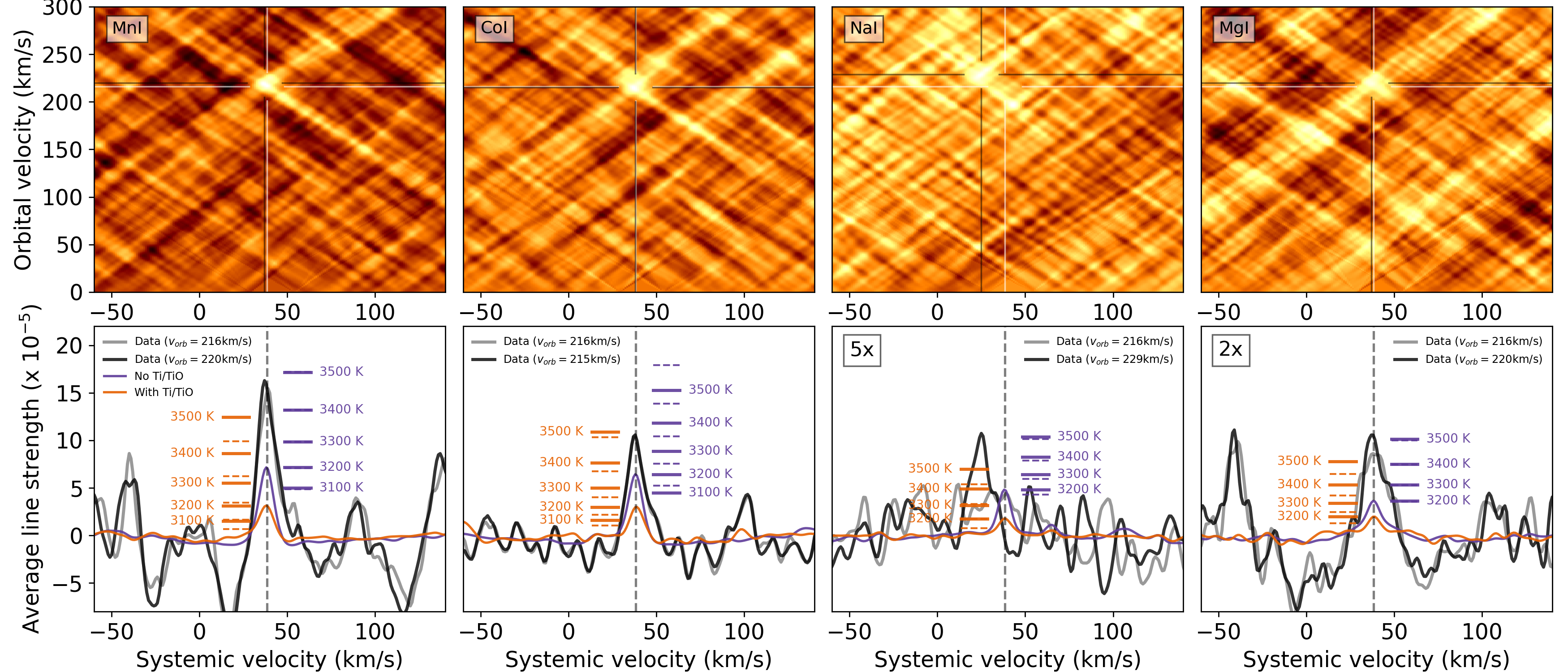}
      \caption{Similar to Fig. \ref{Fig:CCF_results_1}, for Mn, Co, Na and Mg. Mn and Co are detected significantly, while Na and Mg are tentative because of the presence of a strong shift in systemic and orbital velocity in the apparent signal of Na \citep[though see][]{Seidel2023}, and an strong unexplained alias in the CCF of Mg.}
         \label{Fig:CCF_results_3}
  \end{figure*}

We co-added the cross-correlation functions in the planetary rest-frame for each of the time-series, and then combined these with weights equal to the relative detection strength of \ion{Fe}{I} of the injected model with $T=3,200$\,K and $5\times$ solar metallicity (see Table \ref{tab:obs}). This produces the velocity-velocity diagrams that are shown in Fig. \ref{Fig:CCF_results_1}, Fig. \ref{Fig:CCF_results_2_TiTiO} and Fig. \ref{Fig:CCF_results_3} for \ion{Ca}{I}, \ion{Cr}{I}, \ion{Fe}{I}, \ion{Ni}{I}, \ion{Mn}{I}, \ion{Co}{I}, \ion{Ti}{I}, TiO, \ion{V}{I} and VO, as well as tentative signals of \ion{Na}{I} and \ion{Mg}{I}. The cross-correlation functions of all tested species are shown in the appendix in Fig. \ref{Fig:CCF_results_all}, including species for which no detection was made. Based on these results, we report detections of line emission by \alldetectedspecies.

  \begin{figure}
    \includegraphics[width=\columnwidth]{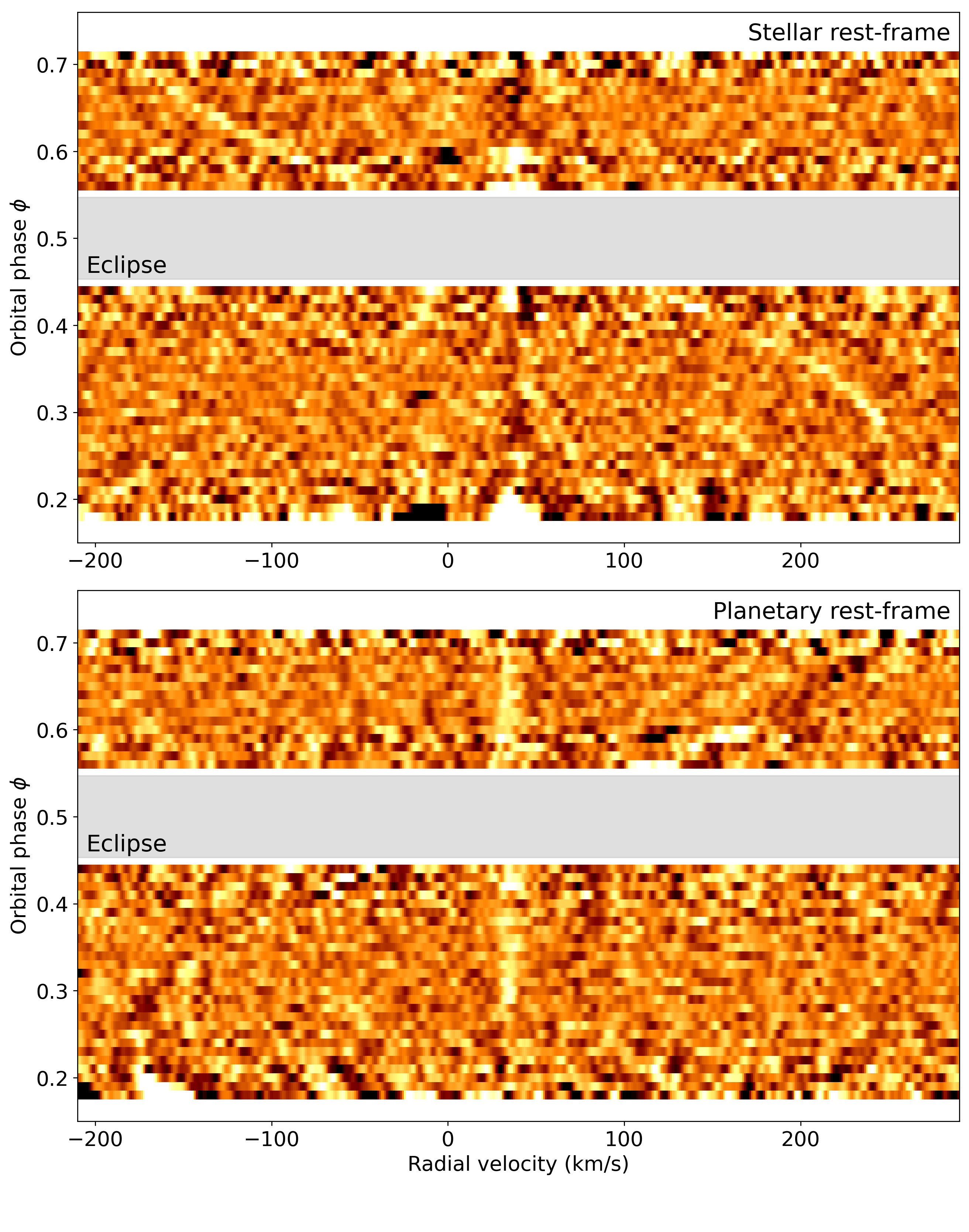}
      \caption{The cross-correlation functions of \ion{Fe}{I} of all 8 time-series binned onto a common phase-grid with steps of 0.01, in both the stellar rest-frame (top panel) and the planetary rest-frame (bottom panel). The trace caused by the emission of \ion{Fe}{I} is clearly visible and increases towards the secondary eclipse. The exposures are averaged with equal weights and uncertainties are propagated accordingly when fitting the model of Eq. \ref{eq:model_ccf}.}
         \label{Fig:trace}
  \end{figure}

Besides co-adding the cross-correlation functions into velocity-velocity diagrams, we rebin and average the cross-correlation functions of the eight epochs onto a common phase grid with a step size of 0.01 in phase, without applying weights. This allows the Doppler-shifted emission of the planet to be visualized as a function of the orbital phase (see Fig. \ref{Fig:trace}), while a lack of weights precludes biasing towards a particular epoch or phase-range (at some cost in signal-to-noise ratio). \ion{Fe}{I} emission clearly traces the radial velocity of the planet both prior to the secondary eclipse and afterwards. The line strength increases towards the secondary eclipse which is expected because most of the hotter sub-stellar point is in view. Using the same phase function as before, we model the two-dimensional emission profile as:

\begin{equation}
    \text{CCF} = A e^{\frac{-(v-v_t)^2}{2 \sigma_w^2}}  \cdot \left( 0.5 + 0.5\cos\left(2 \pi (\phi_t-\theta) -\pi \right) \right)^{\gamma} + C,
\end{equation} \label{eq:model_ccf}

with $v_t = v(t) = v_{\textrm{sys}}+v_{\textrm{orb}} \sin(2 \pi \phi_t) \sin(i)$ the radial velocity of the planet as a function of orbital phase $\phi_t=\phi(t)$. The model has the following free parameters and we define priors for the purpose of carrying out Bayesian inference:
\begin{itemize}
    \item The peak amplitude of the emission line $A \sim \mathcal{U}(-300,3000)$ in ppm.
    \item The systemic velocity $v_{\textrm{sys}} \sim \mathcal{U}(32,44)$ in \kms, centered on the known systemic velocity of \systemicvelocity \kms \citep{Bourrier2020}.
    \item The orbital velocity $v_{\textrm{orb}} \sim \mathcal{U}(210,230)$ in \kms.
    \item The Gaussian line-width $\sigma_w \sim \mathcal{U}(2,10)$ in \kms.
    \item The offset angle of the peak emission $\theta \sim \mathcal{U}(-70,70)$ in degrees, analogous to the hot-spot offset typically fit in broad-band phase-curve observations.
    \item The index of the phase function $\gamma \sim \mathcal{U}(0,6)$
    \item A constant offset $C \sim \mathcal{U}(-3,3)$ in ppm.
\end{itemize}

We sample from these prior distributions and evaluate the likelihood in a Bayesian framework using a No U-Turn Sampler \citep[see][for a review]{Betancourt2017} implemented using \texttt{NumPyro} and \texttt{Jax} \citep{jax2018github,bingham2019pyro,phan2019composable}. Fitting results are summarized in Table \ref{tab:fits}, and the posterior distributions and the best-fit models for all detected species are shown in Fig. \ref{Fig:posteriors}.

\subsection{Velocity traces}
\begin{figure*}
    \includegraphics[width=\textwidth,trim={3cm 0 3cm 1cm},clip]{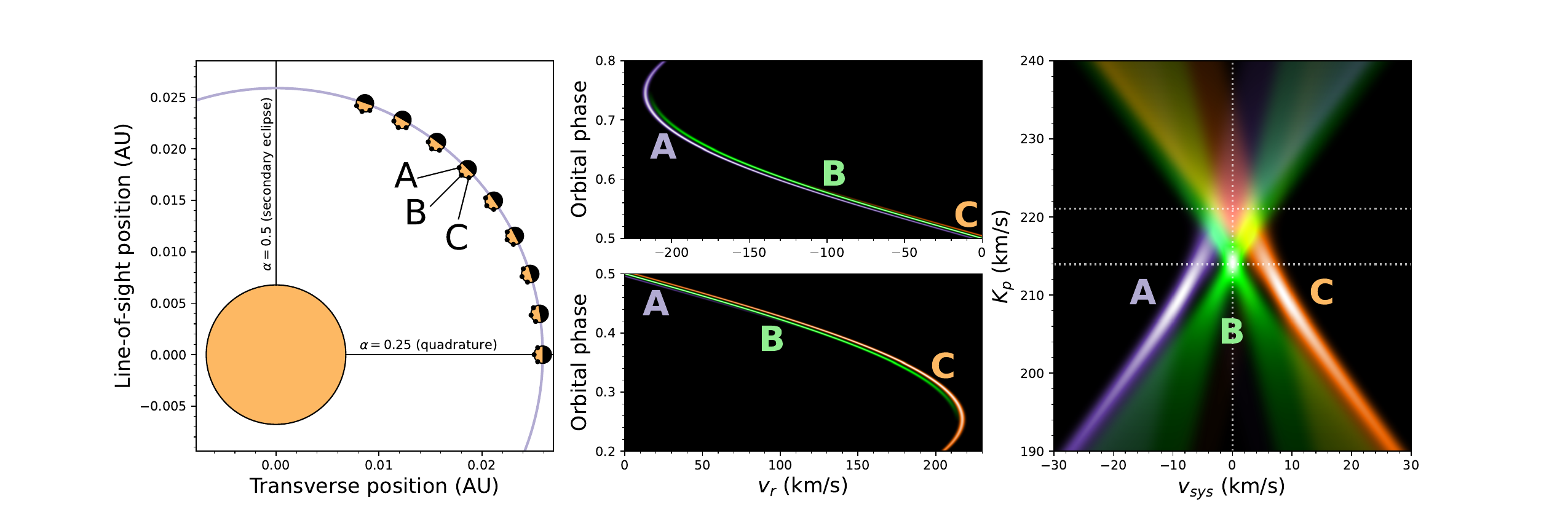}
      \caption{Model of the velocity traces of three points on the day-side near the morning limb (A), sub-stellar point (B) and the evening limb (C). Left panel: Schematic of the WASP-121\,b system to scale, with orbital phases between quadrature and secondary eclipse. The line-of-sight is along the vertical axis. Middle panels: Model of the cross-correlation traces, assuming that line emission originates from each of the three spots in equal strength, with a line-width corresponding to the instrumental resolving power. Planetary rotation causes small differences in the location of each trace. Right panel: Co-added velocity-velocity diagram showing the signal due to each of the three points. White dotted lines indicate the true orbital velocity and the velocity at which the sub-stellar point peaks.}
         \label{Fig:surface}
  \end{figure*}

The parameters that describe the shape and location of the trace ($v_{\textrm{orb}}$, $v_{\textrm{sys}}$, $\gamma$, $\theta$ and $\sigma_w$) are consistent between all species to within $3\sigma$, though variations may exist in particular in the apparent value of $v_{\textrm{orb}}$. This has been observed previously in transit transmission spectra of UHJs, where planetary rotation and dynamical effects cause variations in the peak location of cross-correlation signals in $K_p$-$v_{\textrm{sys}}$ space \citep[e.g.][]{Prinoth2022} as modeled using global circulation models \citep{Wardenier2021,Lee2022}. Without recourse to a GCM, we follow \citet{Pino2022} in producing a simple model of narrow line-emission originating from particular points on the tidally locked surface of WASP-121\,b. The velocity of a point on the equator is assumed to be equal to the (circular) orbital velocity plus synchronous rotation of the planet, leading to the following expression:

\begin{equation}
    v_r(t)|_{\theta} = \frac{2 \pi}{P} \left( a \sin(\phi_t) - R_p \sin(\phi_t-\theta) \right)
\end{equation}

with $R_p$ the radius of the planet, $a$ the semi-major axis of the orbit (which may also be expressed in terms of $P$ and $M_*$), and $\theta$ now equal to the angle between the emitting point and the sub-stellar point. A reduction in orbital velocity may thus be large for inflated, close-in planets (with large values of $\frac{R_p}{a}$). We construct a model of the emission lines of three points on the surface, for $\theta=\pm0.3\pi$ and $\theta=0$, i.e. near the limbs on the day-side and the sub-stellar point -- as a function of radial velocity and orbital phase (i.e. a CCF, see Fig. \ref{Fig:surface}), with equal strength but scaled with a projection factor $\cos{\phi_t - \theta - \pi}$. We then construct a velocity-velocity diagram like before, and note that synchronous rotation of the surface lowers the orbital velocity at which the emission signal maximizes significantly by approximately 10 km/s. Emission from points offset from the sub-stellar point may additionally be significantly blue or red-shifted. We note that the peak of emission for our detected species tend to cluster lie near 216 km/s, approximately 5 km/s below the true orbital velocity of \orbitalvelocity km/s. This could indicate that the majority of emission originates from near the sub-stellar point. The tentative signal of \ion{Na}{I} stands out, with a significantly higher orbital velocity and blue-shift, which may be related to the observation by \citet{Seidel2023} of a high-velocity sodium component. On the other hand, we note that our fit of its trace is poorly converged (see Fig. \ref{Fig:posteriors}), so more observations are needed to confirm the presence of \ion{Na}{I} emission. The effect of planetary rotation may be capable of explaining similar shifts in H$_2$O and CO emission observed in WASP-18\,b by \citet{Brogi2023}, or the seeming orbital eccentricity of KELT-9\,b by \citet{Pino2022}.\\

We additionally detect no significant offset in the phase angle of peak emission for any species, and the model that was injected with an index of $\gamma = 1$ is not significantly discrepant for any of the species. We note that observations of in particular \ion{Fe}{I} emission in ultra-hot Jupiters can be very sensitive to shape-parameters and peak-offsets, and comparisons with predictions from global circulation models as well as observations of similar systems around brighter host stars are warranted. We also acknowledge that in the presence of significant post-processing of the observed spectra (e.g. in the form of filtering algorithms, imperfect removal of telluric lines or stellar lines, or dividing out the time-average especially near quadrature), the planetary line shape can change significantly \citep{Gibson2022} potentially biasing the location of the best-fit parameters and needs to be modeled consistently. However in the present analysis, the most significant filtering was a relatively wide high-pass filter. Most nights cover a large shift in radial velocity (the smallest being 15 \kms in night \# 5) and all telluric lines deeper than 20\% were masked entirely. Besides the decreased orbital velocity due to the planets' rotation, we do not observe anomalous shifts in radial velocity and therefore conclude that atmospheric dynamic effects do not seem to significantly impact our analysis.

\begin{table*}
\caption{Summary of the best-fit parameters obtained by fitting the Doppler-shifted emission trace of each of the detected species according to equation \ref{eq:model_ccf}. These values reflect the median and standard deviations of the marginalized posterior distributions for purposes of clarity. In reality, some posteriors are asymmetric. Refer to Fig. \ref{Fig:posteriors} for the uncertainty intervals in the positive and negative direction separately.}\label{tab:fits}
\centering
\begin{tabular}{cccccccc}
\hline \hline
Element & $A$ & $C$ & $\gamma$ & $\sigma_{w}$ & $\theta$ & $v_{orb}$ & $v_{sys}$ \\
\hline
\ion{Ca}{I} & $149 \pm 40$ & $-0.53 \pm 0.68$ & $3.1 \pm 1.5$ & $6.36 \pm 0.96$ & $4 \pm 12$ & $220.2 \pm 1.7$ & $37.8 \pm 1.1$ \\
\ion{Ti}{I} & $-19 \pm 19$ & $0.05 \pm 0.41$ & $3.7 \pm 1.6$ & $4.8 \pm 2.1$ & $17 \pm 46$ & $221.5 \pm 7.0$ & $35.3 \pm 3.0$ \\
\ion{V}{I} & $92 \pm 17$ & $-0.33 \pm 0.54$ & $0.68 \pm 0.64$ & $3.98 \pm 0.74$ & $23 \pm 23$ & $216.8 \pm 1.1$ & $39.09 \pm 0.78$ \\
\ion{Cr}{I} & $81 \pm 18$ & $-0.37 \pm 0.53$ & $1.15 \pm 0.84$ & $5.12 \pm 0.89$ & $11 \pm 21$ & $214.9 \pm 1.3$ & $39.8 \pm 1.0$ \\
\ion{Mn}{I} & $137 \pm 39$ & $-0.7 \pm 1.3$ & $1.3 \pm 1.2$ & $5.7 \pm 1.5$ & $-7 \pm 27$ & $216.9 \pm 2.5$ & $39.2 \pm 2.1$ \\
\ion{Fe}{I} & $130 \pm 15$ & $-0.47 \pm 0.35$ & $1.37 \pm 0.45$ & $4.21 \pm 0.36$ & $-1.5 \pm 7.6$ & $216.24 \pm 0.51$ & $38.72 \pm 0.37$ \\
\ion{Ni}{I} & $177 \pm 39$ & $-0.6 \pm 1.1$ & $0.42 \pm 0.62$ & $3.12 \pm 0.71$ & $3 \pm 31$ & $214.89 \pm 0.83$ & $38.33 \pm 0.71$ \\
\ion{Co}{I} & $194 \pm 55$ & $-0.49 \pm 0.94$ & $4.5 \pm 1.1$ & $5.3 \pm 1.4$ & $3.1 \pm 7.0$ & $217.8 \pm 2.2$ & $39.5 \pm 1.3$ \\
\hline
\end{tabular}
\end{table*}



\subsection{Model comparison}
The cross-correlation signals in Figs. \ref{Fig:CCF_results_1} and \ref{Fig:CCF_results_2_TiTiO} are compared with the strengths of models with and without titanium, for temperatures ranging from 3,100\,K to 3,500\,K and for two values of the metallicity of 5 and 10 times solar (0.7 and 1.0 dex). As expected, the line strength increases strongly with the assumed temperature difference across the inversion. Removing Ti from the atmosphere causes the removal of TiO. Because TiO absorbs strongly and thereby masks lines of other species, models with Ti require higher temperatures to match the observed signals of other atoms. For example, the signal of \ion{Ca}{I} is well reproduced by a model with a metallicity of $5\times$ solar and a peak temperature of 3,400\,K if Ti/TiO are present, but only 3,200\,K is needed without Ti/TiO. This underscores the importance of model completeness in exoplanet spectroscopy: Even molecules that are not individually detected or detectable may produce significant (pseudo-continuum) absorption that affects the relative line strengths of other species that are observed. The success of spectroscopic model inference depends on the completeness of the set of opacities that are included in the model. In the case of WASP-121\,b previously, missing opacity also led to an apparently anomalous spectral slope in the transmission spectrum that is reproduced well when metal line absorption is included \citep{Hoeijmakers2020}.\\

Fig. \ref{Fig:CCF_results_2_TiTiO} shows non-detections of both \ion{Ti}{I} and TiO, a robust detection of \ion{V}{I} and a non-detection of VO, which is predicted to be too faint to detect with models at these temperatures unless the metallicity is significantly higher. From these comparisons, we conclude that models that contain Ti/TiO are not capable of reproducing the data for two reasons:

\begin{enumerate}
    \item TiO is strongly ruled out, even for the model with the lowest expected TiO line strength (3,100\,K and 10$\times$ solar metallicity).
    \item To explain the line strengths of the other species, a peak temperature of approximately 3,400\,K is needed for models with Ti/TiO. However, at such a temperature, \ion{Ti}{I} would be detected very strongly.
\end{enumerate}

Furthermore, based on the model comparison, we predict that with additional observations, other species may be detectable if they are not depleted similarly to \ion{Ti}{I}, including \ion{Sc}{I} and \ion{Y}{I} (see Fig. \ref{Fig:CCF_results_all}). We also note that generally, ions that are very strongly observed in transmission are essentially undetectable in emission in this planet and this is expected given that the emitted flux decreases sharply towards shorter wavelengths, and the fact that ions mainly exist at much lower altitudes to which emission spectroscopy is inherently less sensitive.

\subsection{Depletion of the Ti inventory}
Ti is not detected in emission in this study, nor in previous observations of the transmission spectrum \citep{Hoeijmakers2020,Merritt2021,Gibson2022} even though other absorbers are, in particular V. The injected models, which reproduce detections of other species within a factor of order unity, predict that Ti should have been significantly detected. Similarly, TiO is expected to be detected strongly assuming that the line-list \citep{McKemmish2019} is sufficiently accurate. The recent detection of TiO in the transmission spectrum of WASP-189\,b indicates that it is \citep{Prinoth2022}. A recent study of global circulation models of this planet using non-elevated metallicity predicts that the thermal inversion occurs at slightly higher pressures (100 to 10 mbar), and that under such circumstances TiO should still be recoverable from this data, even if Ti may not be \citep{Lee2022}. Note that our models assume solar elemental ratios, while the abundance of Ti in the host star WASP-121 is significantly greater than that of V \citep{Polanski2022}, making a non-detection of Ti relative to V even more robust.

The apparent absence of emission by Ti and TiO suggests that the abundances of Ti and Ti-bearing molecules are reduced compared to what is expected from stellar elemental abundance ratios and chemical equilibrium. Assuming that there is no bulk depletion of Ti in \WASP, there is a physical mechanism that causes depletion of Ti from the hot day-side as well as the terminator regions. We hypothesize that this depletion is caused by condensation of Ti, predicted to condense as CaTiO$_3$ (perovskite) at one of the highest temperatures of all atoms \citep{Lodders2002}. Recent observations of the phase-curve of \WASP with HST \citet{Evans2022} imply that the night-side temperature profile indeed lies below the condensation curve of perovskite, meaning that significant condensation of perovskite should occur on the night-side. However, on the hot day-side with a temperature over 3,000\,K, condensation is not expected to play a role. Instead, we infer that night-side condensation of Ti causes it to be cold-trapped \citep{Spiegel2009}, unable to recirculate back to regions of the atmosphere probed by emission and transmission spectroscopy. The night-side cold-trap thus causes titanium to be depleted globally, including in regions where the temperature is far above its condensation curve. Although titanium initially condenses with Ca to form \perovskite, this is not expected to significantly alter the atmospheric Ca abundance because it has a significantly greater elemental abundance \citep{Asplund2009}.\\

\begin{figure*}
    \includegraphics[width=\textwidth,trim={0cm 0cm 0cm 0cm},clip]{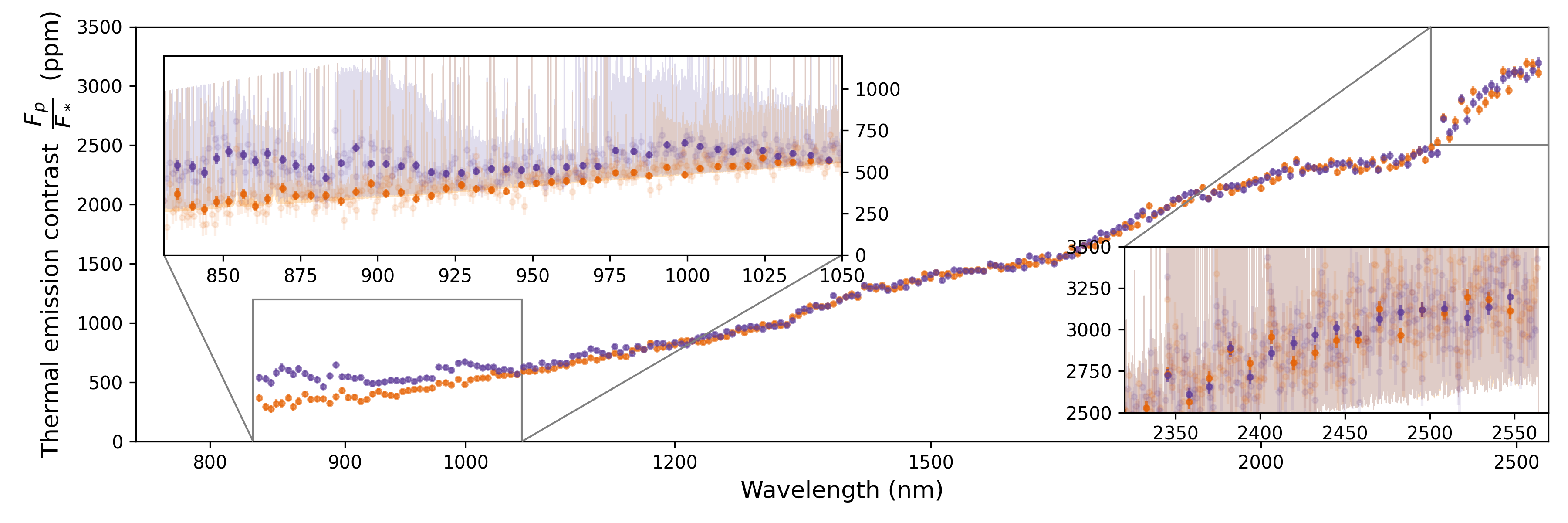}
      \caption{Simulated emission spectra of the day-side as observed with JWST/NIRISS SOSS, using Pandexo \citep{Batalha2017}. Solid lines: high-resolution models assuming a temperature inversion up to 3200 K (see Fig. \ref{Fig:TP}) at a metallicity of 5 $\times$ solar (solid lines). Transparent dots: Pandexo simulation at the native instrument resolution. Solid dots: binned to a resolving power of 100. Purple: Model with Ti/TiO. Orange: Model without Ti/TiO. If TiO is indeed absent from the day-side emission spectrum of WASP-121\,b, TiO bands will be strongly ruled out at the shortest wavelengths of NIRISS, while the VO band strength near 1 $\mu$m will be hard to discern. Note that the water band at $1.4 \mu$m is barely visible by eye due to the scale of the axes, but the increased contrast from 1000 to 1500 ppm between $1.3$ and $1.6 \mu$m is consistent with the WFC3 data \citep[see Fig. 2][]{Evans2017}. }
         \label{Fig:JWST}
  \end{figure*}

The present observations do not constrain where, or how the titanium is trapped. It may be that titanium-bearing condensates are prevented from recirculating back to the day-side due to inefficient advection (horizontal mixing), although \citep{Seidel2023} find evidence for strong day-to-night flows. Alternatively, titanium-bearing condensates may indeed recirculate back to the day-side but are confined to high pressures, at altitudes below the thermal inversion on the day-side, (i.e. below the mbar to $\mu$bar pressures probed in emission and transmission). We believe that this scenario is currently not supported by global circulation models, which predict strong vertical mixing \citep{Parmentier2013,Menou2019} that increases strongly with planet equilibrium temperature \citep{Komacek2019}. However, regardless of the location at which the titanium is trapped, we derive several important implications from these observations:

\begin{enumerate}
    \item High-resolution emission spectroscopy of ultra-hot Jupiter atmospheres can be used to impose limits on the efficiency with which condensates are remixed to the high-temperature day-side by global circulation, or if it is remixed horizontally, the efficiency by which it is mixed vertically across the temperature inversion (i.e. $K_{zz}$). In the latter case, these observations demonstrate that the atmospheres of ultra-hot Jupiters cannot be assumed to be well mixed on the day-side.
    \item The process of condensation and limited remixing observed in this study is specific to the Ti inventory, and not e.g. Fe, V, Ni or Cr. Therefore, we hypothesize that the temperature profile on much of the night-side lies between the condensation curves of CaTiO$_3$ and VO. This is fully consistent with the night-side temperature profile derived from spectroscopic phase-curve observations recently published by \citet{Evans2022}.
    \item In this formulation of the cold-trap hypothesis, an element is depleted from the gas-phase on a global scale if the night-side temperature profile is below a threshold and if re-circulation or vertical mixing is inefficient. This implies that planets with slight differences in equilibrium temperature and/or global circulation properties can have drastically different atmospheric compositions, with one or multiple species depleted compared to less volatile species. For example, on planets with slightly lower night-side temperatures than \WASP both Ti and V may be removed from the gas-phase globally. Conversely, a slightly hotter planet may have both titanium and vanadium-bearing species in the gas-phase. We therefore predict that there exists no smooth continuum in the atmospheric composition (terminator nor day-side) of hot Jupiters as a function of equilibrium temperature. Instead, we expect that there are sharp chemical transitions with a complex dependency on both equilibrium temperature and global circulation. 
    \item At solar metallicity, aluminum condenses at higher temperatures than titanium in the form of spinel (MgAl$_2$O$_4$)\footnote{Although aluminum may condense along with magnesium in the form of spinel, this should not significantly deplete the magnesium reservoir because magnesium is much more abundant than aluminum \citep{Asplund2009}.} or corundrum (Al$_2$O$_3$) \citep{Lodders2002}. If our formulation of the cold-trap effect is correct, this means that aluminum should be expected to be cold-trapped together with titanium. This would explain the absence of \ion{Al}{I} absorption from the HST/STIS spectrum \citep{Evans2018}, rather than photo-ionization as hypothesized by us earlier \citep{Hoeijmakers2020}. This also means that bands of AlO should be generally unobservable in planets cooler than WASP-121\,b, e.g. in WASP-43\,b \citep{Chubb2022}.
    \item As metals and metal-oxides are expected to provide important sources of heating of the upper atmospheres of hot Jupiters \citep{Lothringer2018}, we predict that the temperature structures of the day- and night-sides should be strongly coupled. A consequence is that the atmospheric opacity computed as part of 3D global circulation models need to be computed self-consistently, where the opacity on the day-side is dependent on the minimum night-side temperature and the effects of condensation there. Condensation physics and transport of condensate particles need to be built into GCMs to explain the observed properties of ultra-hot Jupiters.
    \item Simpler 1D models of day-side or terminator temperature profiles also need to take into account non-local condensation in order to be self-consistent. The effect of cold-trapping can be parameterized via reduction of the abundances of certain metals and molecules, but this would need to take into account the condensation sequence: Depletion of e.g. vanadium without depletion of titanium would not be self-consistent.
    \item The fact that metals may be depleted globally due to condensation effects challenges notions of metallicity and elemental abundance ratios to describe the bulk compositions of hot Jupiters. In all likelihood, the bulk titanium abundance of \WASP is not anomalously low even though observations of the day-side and transmission spectra indicate depletion. Any theories that would rely on the bulk titanium abundance \citep[e.g. theories about formation history, see e.g.][]{Lothringer2021} would be poorly constrained. For species that are sensitive to cold-trapping, apparent measurements of elemental abundances or abundance ratios cannot be translated into bulk abundances without taking into account depletion via non-local condensation \citep[this was also explored by ][]{Pelletier2021}. Direct measurements of bulk metal abundances from emission or transmission spectra will only be possible if the planet is sufficiently hot on the night-side to prevent cold-trapping, or if models accurately include condensation chemistry and global transport.
    \item We have simulated the emission spectrum as it is expected to be observed with NIRISS SOSS (see Fig. \ref{Fig:JWST}) as part of GTO program \#1201 (P.I. Lafreniere). These observations will confirm the presence or absence of TiO bands at short wavelengths with very high confidence, but will not be very sensitive to VO as its (integrated) absorption bands are weaker.
\end{enumerate}

\subsection{Ti/TiO as tracers of atmospheric structure and chemistry}
The equilibrium-chemistry abundance profiles of Ti and TiO are shown in Fig. \ref{Fig:Chemistry}. We observe that because of the large temperature difference across the inversion, the predicted abundance of TiO drops by orders of magnitude due to dissociation, sharply increasing the abundance of Ti. As both Ti and TiO have rich optical emission spectra, spectra at optical wavelengths are highly sensitive to the presence of titanium in the case of an inverted atmosphere. We predict that this strong temperature dependence makes the Ti/TiO pair an especially sensitive probe of the T-P profile of ultra-hot Jupiter atmospheres that can be used in high-resolution retrieval analyses \citep{Brogi2019,Gibson2022}, as long as the planet is hot enough to avoid cold-trapping \citep[e.g. WASP-189\,b for which Ti has been detected.][]{Prinoth2022}. The profile of V shows a similar behavior but is less pronounced because the overall abundance of VO is smaller. We also note from Fig. \ref{Fig:Chemistry} that the abundances of Ti and V depend relatively weakly on the elemental Ti and V ratios at temperatures around 2,500\,K. An increase in the metallicity from $5\times$ to $10 \times$ solar increases the abundance of V from approximately 3.5 ppb to 4.5 ppb. Increasing these elemental ratios instead manifests itself mostly as an increase in the abundances of TiO and VO \citep[see also the discussion in ][]{Hoeijmakers2020}. As can be seen in Fig. \ref{Fig:CCF_results_2_TiTiO}, this translates to a strong correlation between the VO emission strength and metallicity.\\

  \begin{figure}
    \includegraphics[width=\columnwidth]{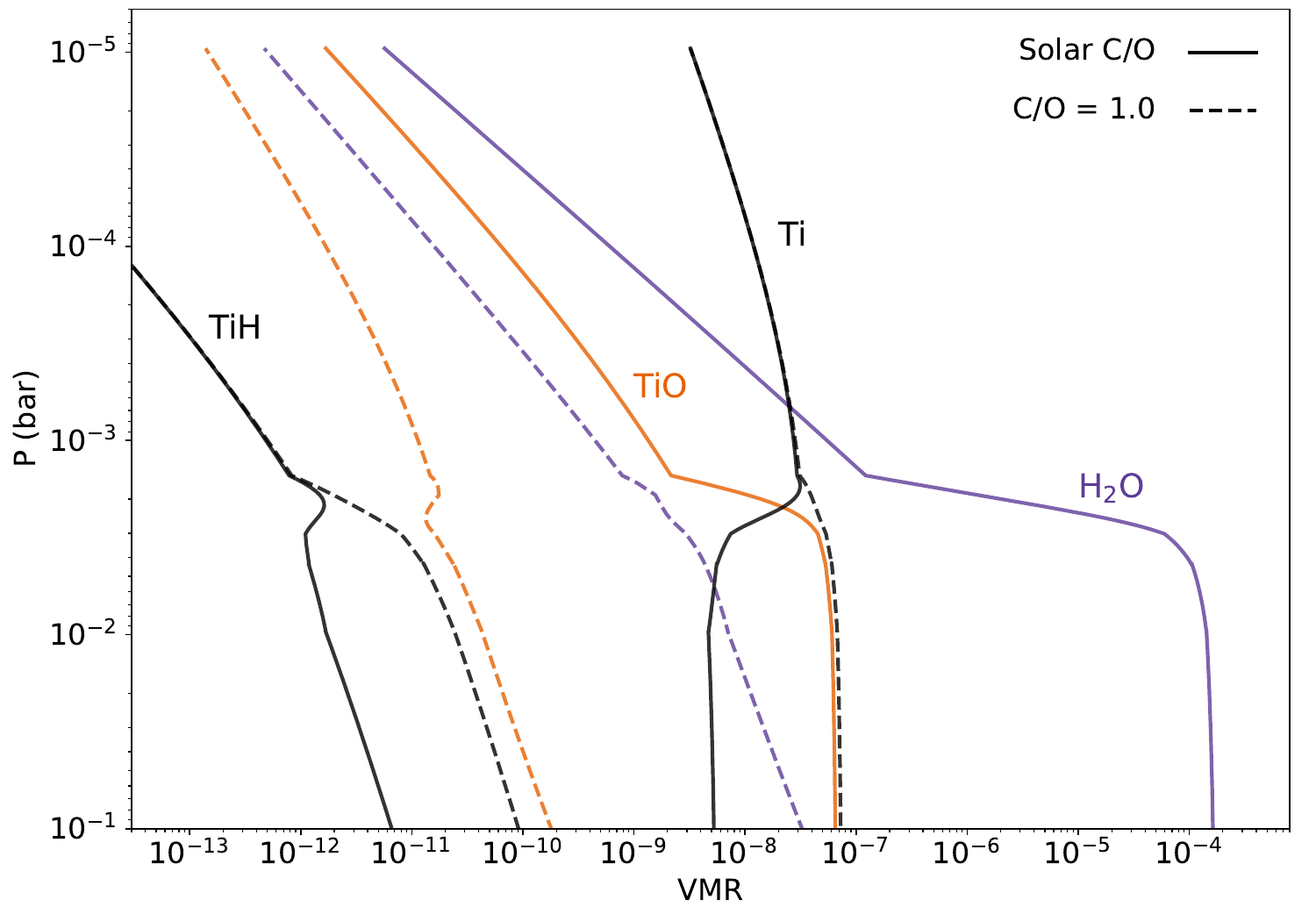}
      \caption{Equilibrium-chemistry solution of the atmosphere of WASP-121\,b assuming a metallicity of $5\times$ solar, a temperature inversion with a maximum temperature of 3200 K and a low or a high C/O ratio. The C/O ratio is varied by modifying the abundance of carbon while leaving oxygen fixed.}
         \label{Fig:CO}
  \end{figure}

\noindent By varying the abundance of carbon, we observe that the abundance of TiO is a strong function of C/O ratio, consistent with the argument originally formulated by \citet{Madhu2011} in the context of TiO/VO-driven thermal inversions, and recent work by \citet{Tsai2021}. In the case of a C/O ratio of 1.0, the TiO abundance drops by three orders of magnitude compared to solar C/O, increasing the abundance of Ti, as well as TiH (see Fig. \ref{Fig:CO}). However an elevated C/O ratio cannot explain our reported absence of TiO in the transmission and emission spectra of WASP-121\,b, as this would be inconsistent with the strong detection of water emission, which suggests an elevated oxygen abundance (used as a proxy for metallicity) \citep{Evans2022}. Nevertheless, Fig. \ref{Fig:CO} implies that the abundance ratio of a metal to its oxide (e.g. Ti/TiO or V/VO) can generally be used to constrain the C/O ratio using spectra at optical wavelengths alone.

\section{Conclusions}
In this paper, we present observations of line emission by metals in the day-side of \WASP using the high-resolution ESPRESSO spectrograph. We report strong detections of \alldetectedspecies, and note an absence of signals by Ti and TiO, indicating that these species are depleted from the day-side atmosphere. This is consistent with earlier observations of the transmission spectrum that also failed to detect these species \citep{Hoeijmakers2020,Merritt2021,Gibson2022}. We infer that titanium condenses on the cold night-side due to the relatively high condensation temperature of perovskite and is unable to circulate back to high altitudes on the day-side and terminator regions, and is hence cold-trapped \citep{Spiegel2009}. We conclude that the chemistry of ultra-hot Jupiters does not lie on a smooth continuum with equilibrium temperature, but instead features strong transitions. Species may deplete globally from the gas phase with small changes in equilibrium temperature, depending on the lowest night-side temperature and the efficiency of advection and vertical mixing. Consequently, we expect that Ti, TiO and also aluminum-bearers should be undetectable for any planet cooler than \WASP, and that the transition above which Ti/TiO become detectable lies between the equilibrium temperatures of \WASP ($\planetteq$) and WASP-189\,b ($2,641 \pm 34 K$) \citep{Prinoth2022}. This causes complex coupling between the day-side and night-side temperature profiles, as metals and metal oxides are strong sources of stratospheric heating on the day-side. Although elemental abundance ratios can be measured from high-resolution spectra \citep{Gibson2022}, the interpretation of such measurements is dependent on our understanding of condensation chemistry on the night-side, which is hard to observe directly. Similarly, although the emission line strength is strongly dependent on the temperature-pressure profile, a temperature determination depends significantly on the presence or absence of masking TiO bands -- a symptom of the more general question of model completeness in exoplanet spectroscopy. We have also found that the apparent orbital velocity of most detected species is significantly smaller than the true orbital velocity, by approximately 5 km/s, and we attribute this to the effect of planetary rotation. This effect implies that unless the distribution of line emission on the day-side is known (e.g. from spectroscopic eclipse mapping observations) or the equatorial rotation velocity is small, high-resolution day-side observations do not provide reliable measurements of the planet orbital velocity and by extension, the dynamical mass of the star. Instead, accurate knowledge of the stellar mass is required in order to obtain reliable insight into atmospheric dynamics as probed via high-resolution cross-correlation spectroscopy.


%

\begin{acknowledgements}
This study is based on observations collected at the European Southern Observatory under ESO program(s) 0105.C-0591 \& 0106.C-0737. This research has made use of the services of the ESO Science Archive Facility.
E.K.H. Lee is supported by the SNSF Ambizione Fellowship grant (\#193448). B.P. \& H.J.H.  acknowledge partial financial support from The Fund of the Walter Gyllenberg Foundation. R. A. is a Trottier Postdoctoral Fellow and acknowledges support from the Trottier Family Foundation. This work was supported in part through a grant from the Fonds de Recherche du Qu\'ebec - Nature et Technologies (FRQNT). This work was funded by the Institut Trottier de Recherche sur les Exoplan\`etes (iREx). JLB acknowledges funding from the European Research Council (ERC) under the European Union’s Horizon 2020 research and innovation program under grant agreement No 805445. We thank S. Pelletier for useful discussions and input.
\end{acknowledgements}

%
%

\bibliographystyle{aa}
\bibliography{standard.bib,refs.bib}

\begin{appendix} 

\section{Overview of all cross-correlation functions, chemistry and posterior distributions}\label{app:posteriors}
\begin{figure*}[ht!]
\includegraphics[width=\textwidth]{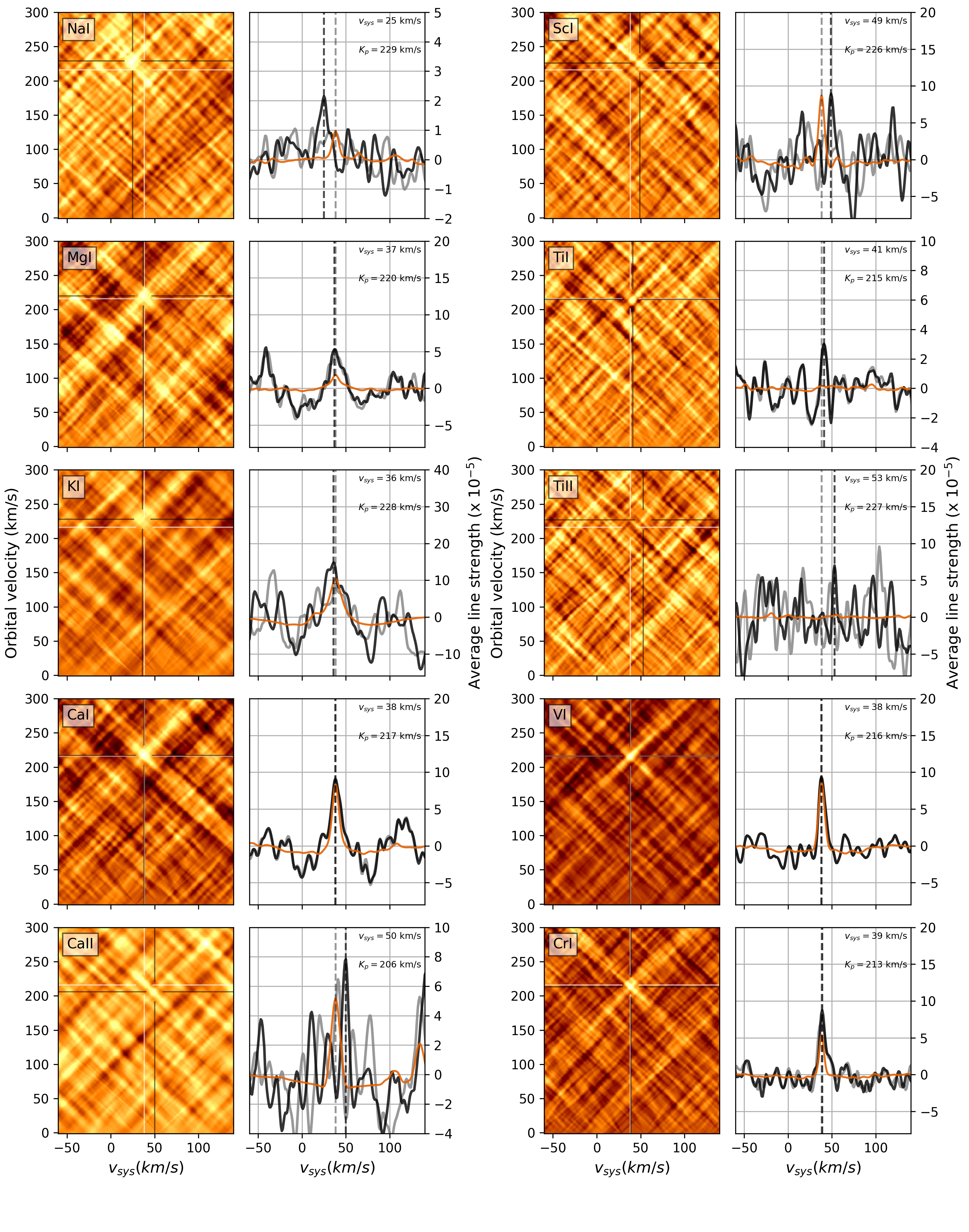}
\caption{Collection of the cross-correlation functions of all species searched for in this study, similar to Fig. \ref{Fig:CCF_results_1}. The cross-correlation functions are organized in two pairs of columns, with the KpVsys diagram on the left, and the one-dimensioanl CCF extracted at an orbital velocity of 216.0 km/s (gray lines) or at the orbital velocity at which the CCF peaks (black lines), as in Fig. \ref{Fig:CCF_results_1}. The values in the top-left corner are the peak position. The orange lines indicate the expected signal of a model spectrum without Ti/TiO and a temperature inversion from 2500 K to 3200 K.}
\label{Fig:CCF_results_all}
\end{figure*}
\begin{figure*}[ht!]
\includegraphics[width=\textwidth]{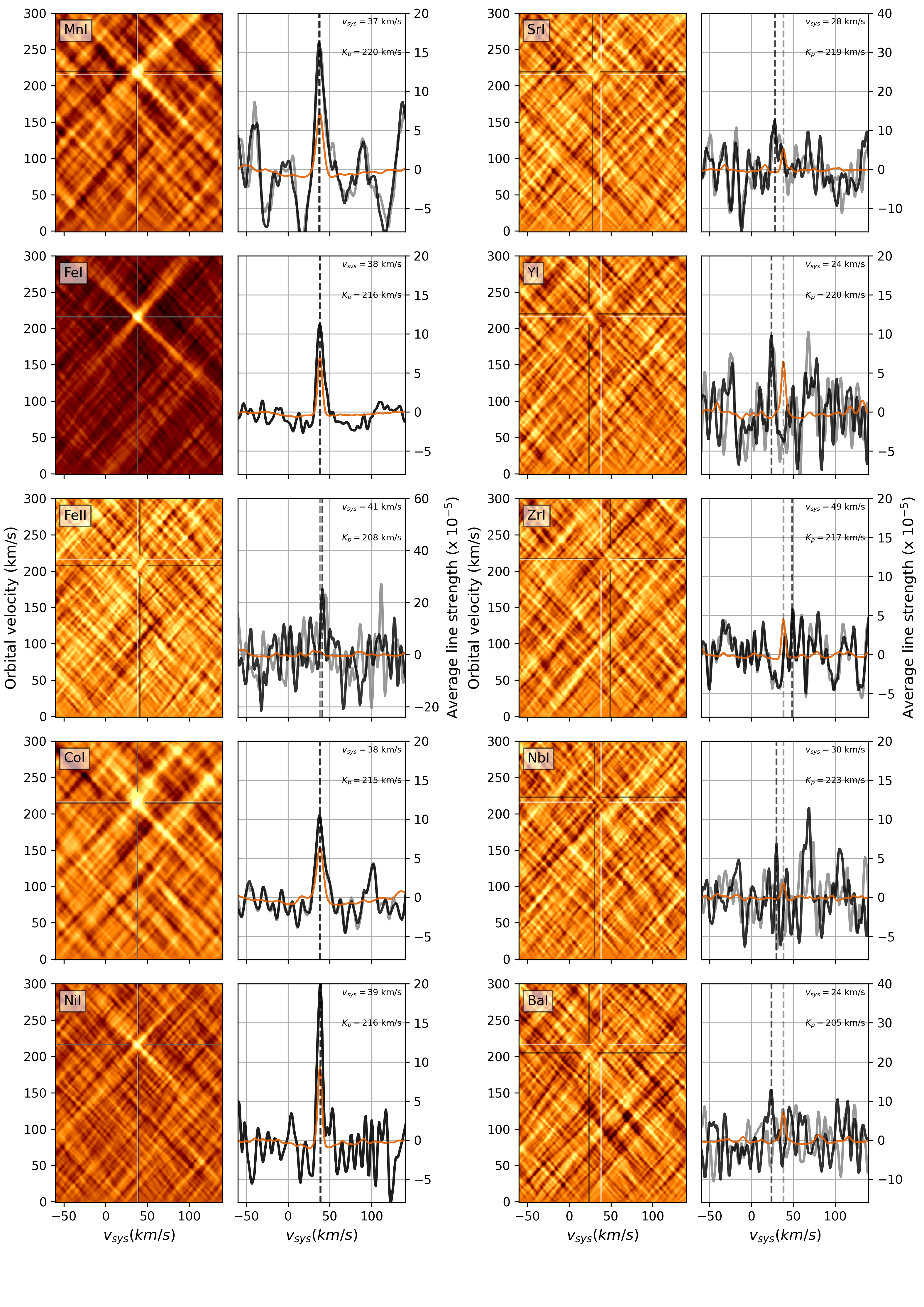}
\caption{Same as Fig. \ref{Fig:CCF_results_all}}
\end{figure*}

\begin{figure*}[ht!]
\includegraphics[width=\textwidth]{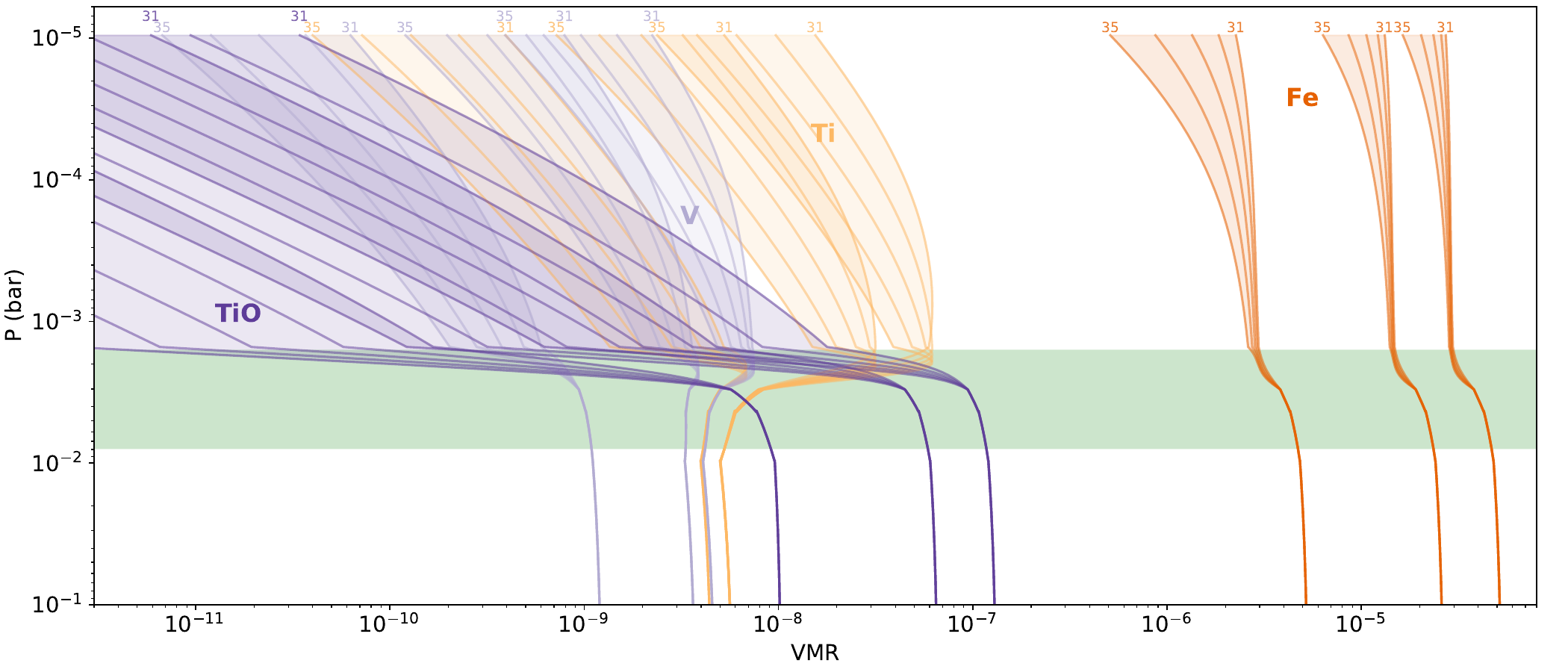}
\caption{The abundances profiles of Fe, Ti, V and TiO, for the different T-P profiles described in Section \ref{sec:models}. Each species is plotted for three different values of the metallicity (increasing from left to right as $1 \times$, $5 \times$ and $10 \times$ solar) and for five values of the isothermal peak temperature (3100 K to 3500 K), indicated by the double digit labels at the top. The shaded horizontal region indicates the range of pressures over which the HST data analyzed by \citet{Evans2022} is sensitive, and where the temperature inversion takes place.}
\label{Fig:Chemistry}
\end{figure*}

\begin{figure*}
\includegraphics[width=\textwidth]{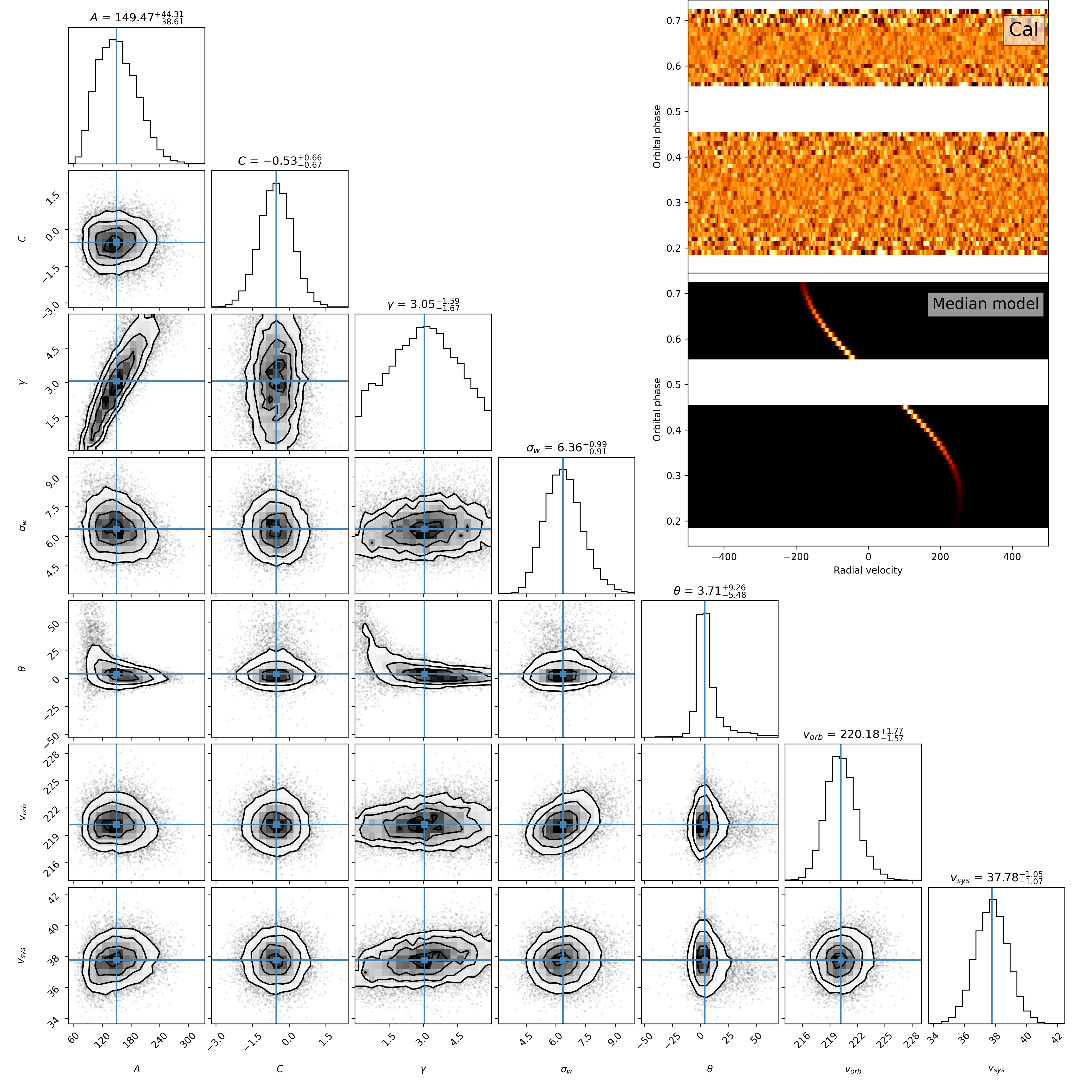}
\caption{Posterior distributions of the model parameters of equation \ref{eq:model_ccf} for \ion{Ca}{I}. The top-right panels show the CCF of all eight epochs combined binned to a common phase grid (top), and the model with parameters equal to the median of their posterior (bottom panel). For \ion{Ca}{I} the signal is poorly visible by eye in the CCF, but for \ion{Cr}{I} and \ion{Fe}{I} it can be discerned (see below). The fit of \ion{Na}{I} is poorly converged and consequently we classify the \ion{Na}{I} signal as tentative.}
\label{Fig:posteriors}
\end{figure*}

\begin{figure*}
\includegraphics[width=\textwidth]{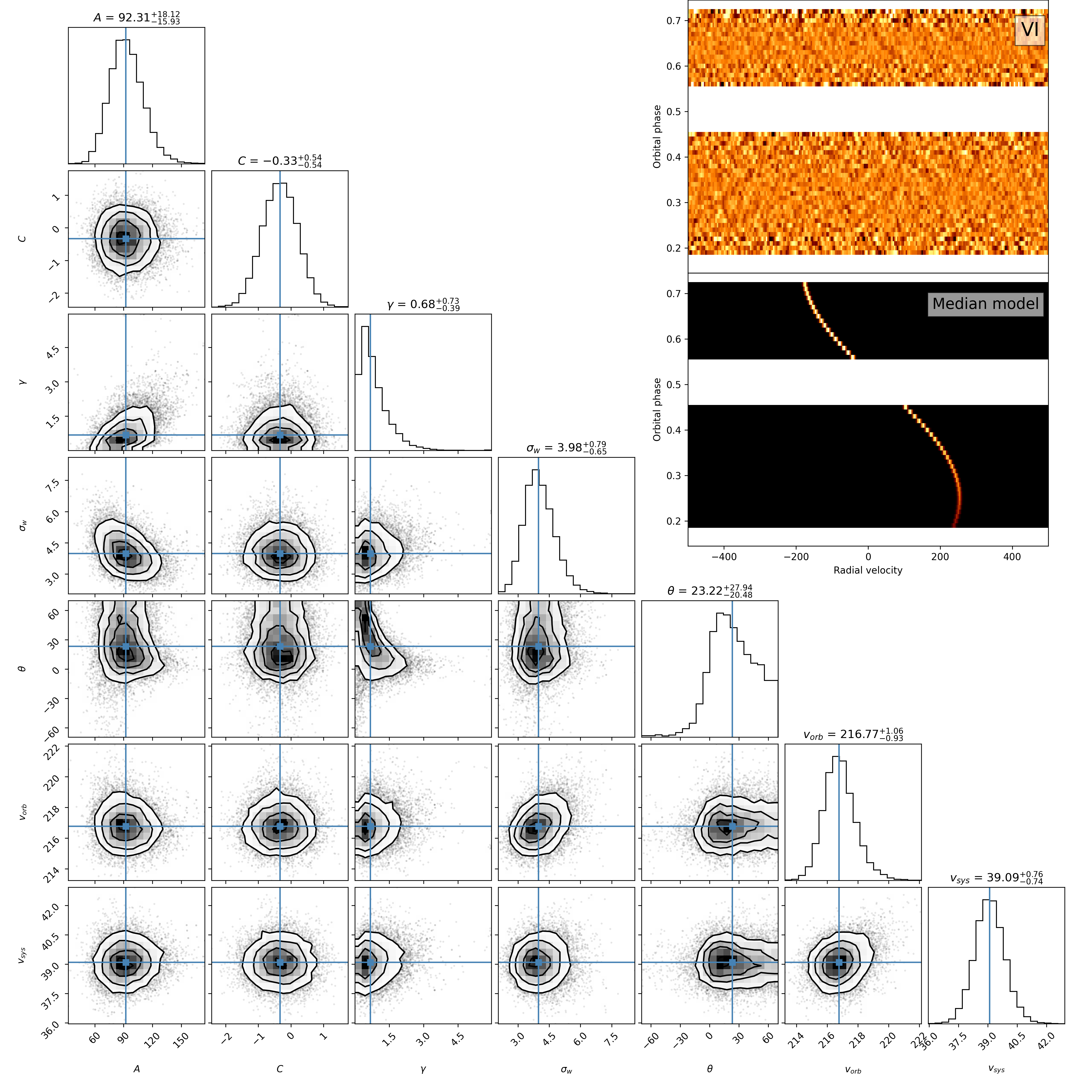}
\caption{Same as Fig. \ref{Fig:posteriors}, for \ion{V}{I}.}
\end{figure*}

\begin{figure*}
\includegraphics[width=\textwidth]{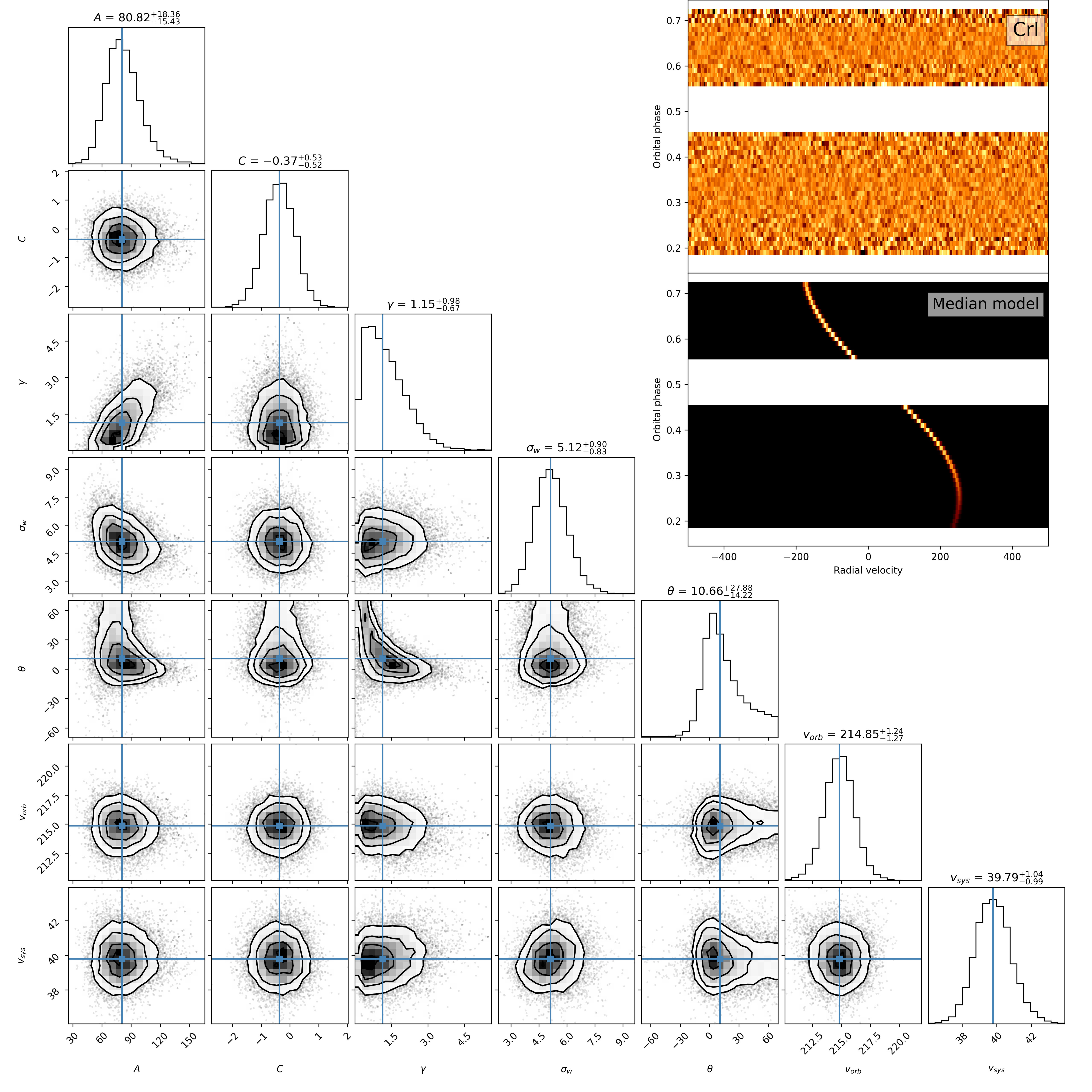}
\caption{Same as Fig. \ref{Fig:posteriors}, for \ion{Cr}{I}.}
\end{figure*}

\begin{figure*}
\includegraphics[width=\textwidth]{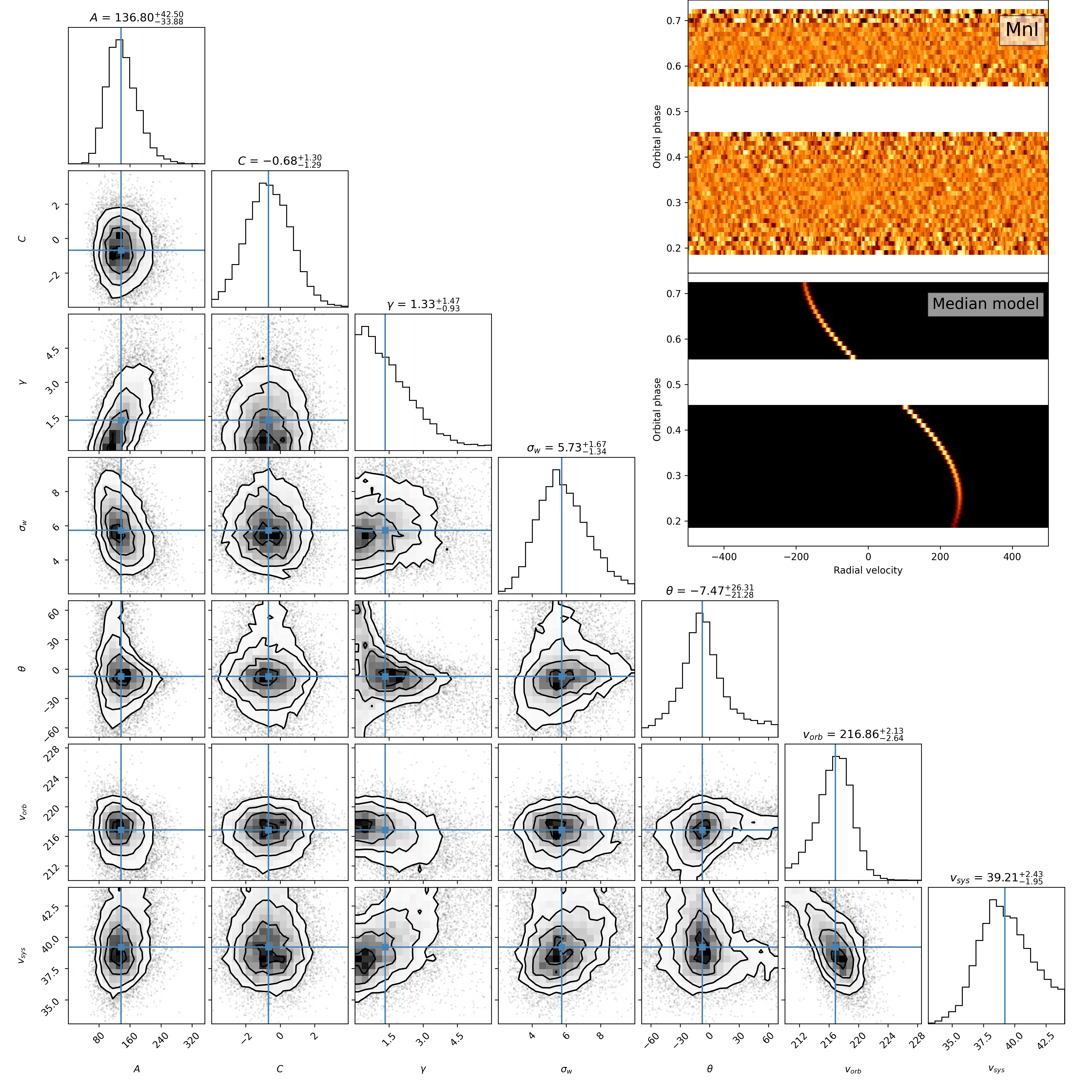}
\caption{Same as Fig. \ref{Fig:posteriors}, for \ion{Mn}{I}.}
\end{figure*}

\begin{figure*}
\includegraphics[width=\textwidth]{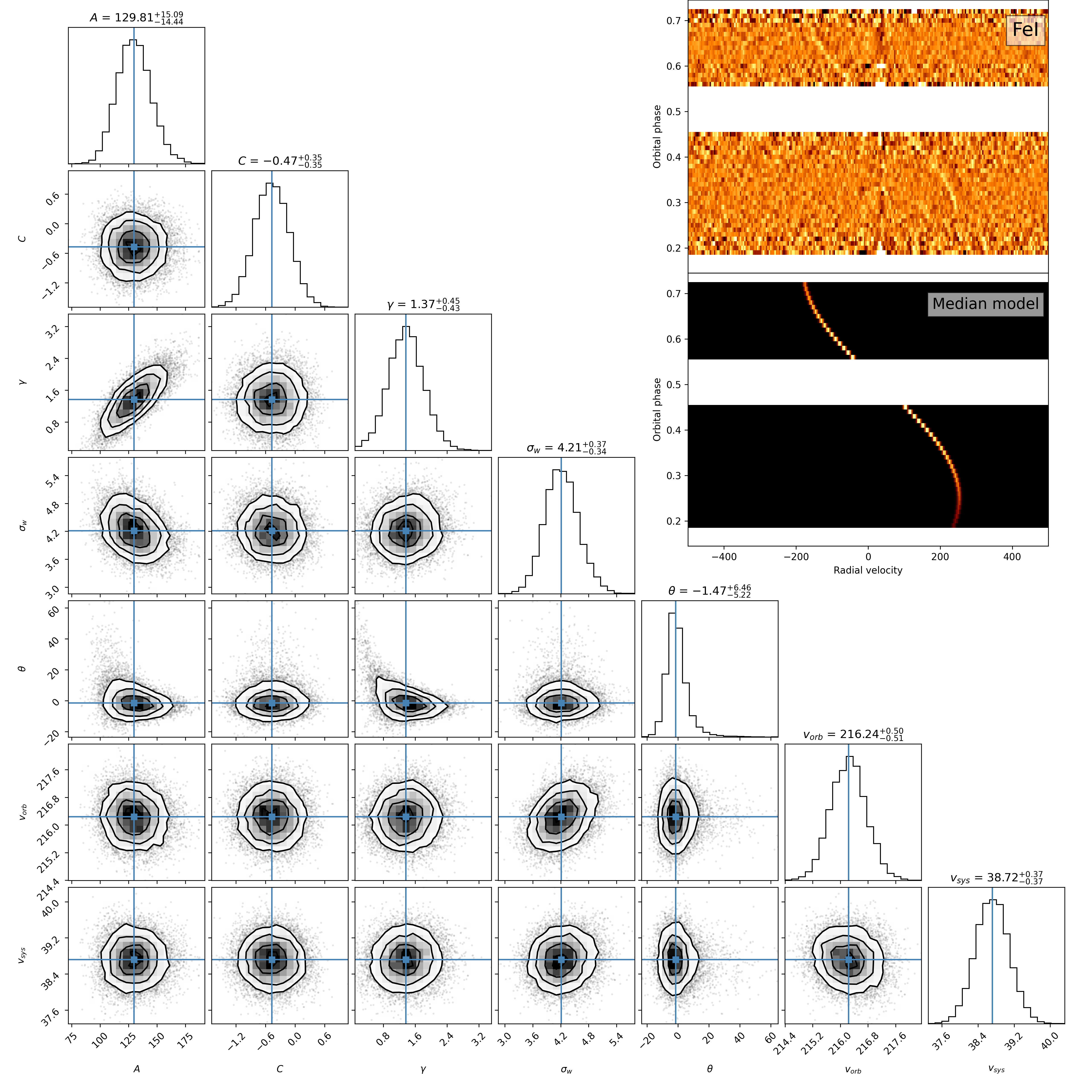}
\caption{Same as Fig. \ref{Fig:posteriors}, for \ion{Fe}{I}.}
\end{figure*}

\begin{figure*}
\includegraphics[width=\textwidth]{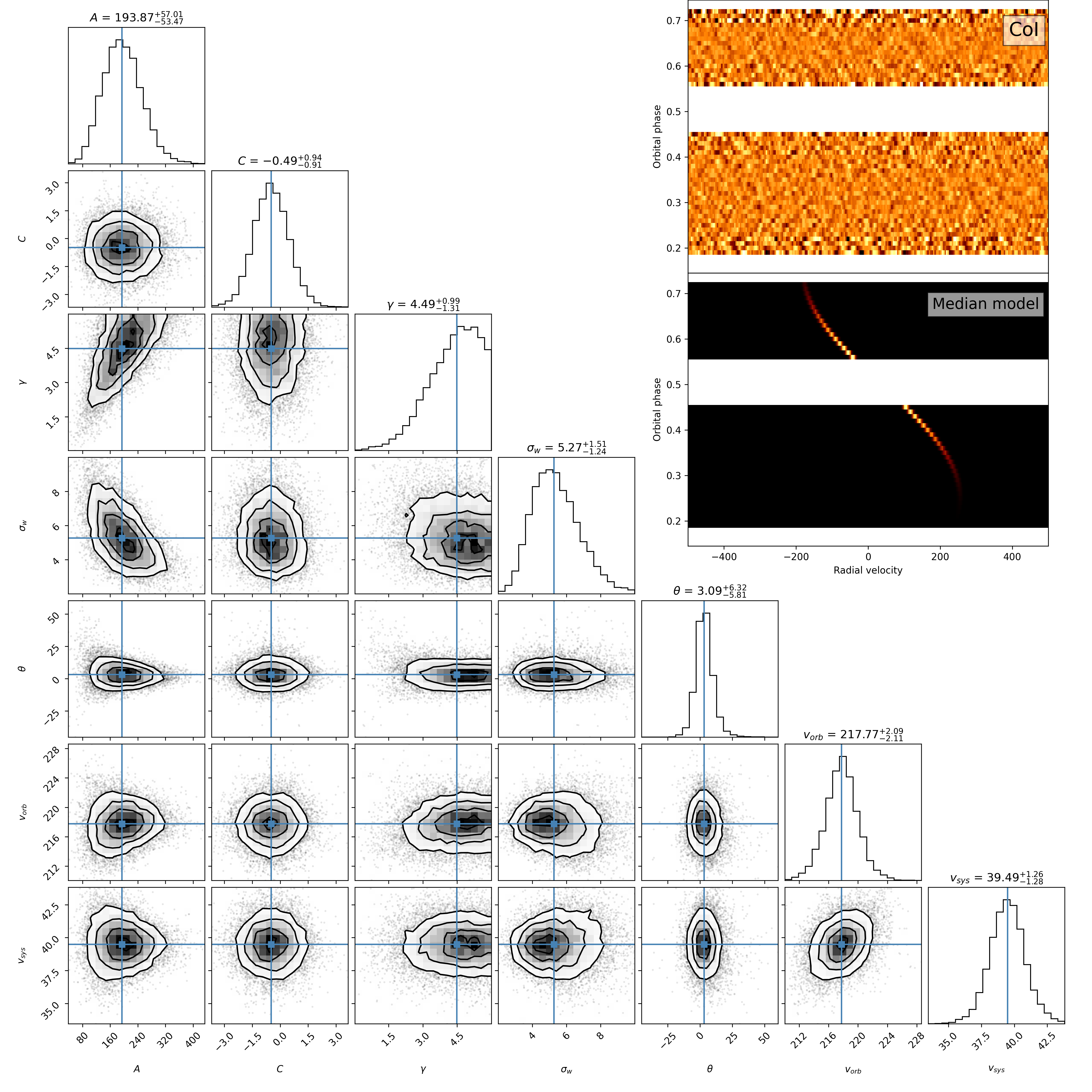}
\caption{Same as Fig. \ref{Fig:posteriors}, for \ion{Co}{I}.}
\end{figure*}

\begin{figure*}
\includegraphics[width=\textwidth]{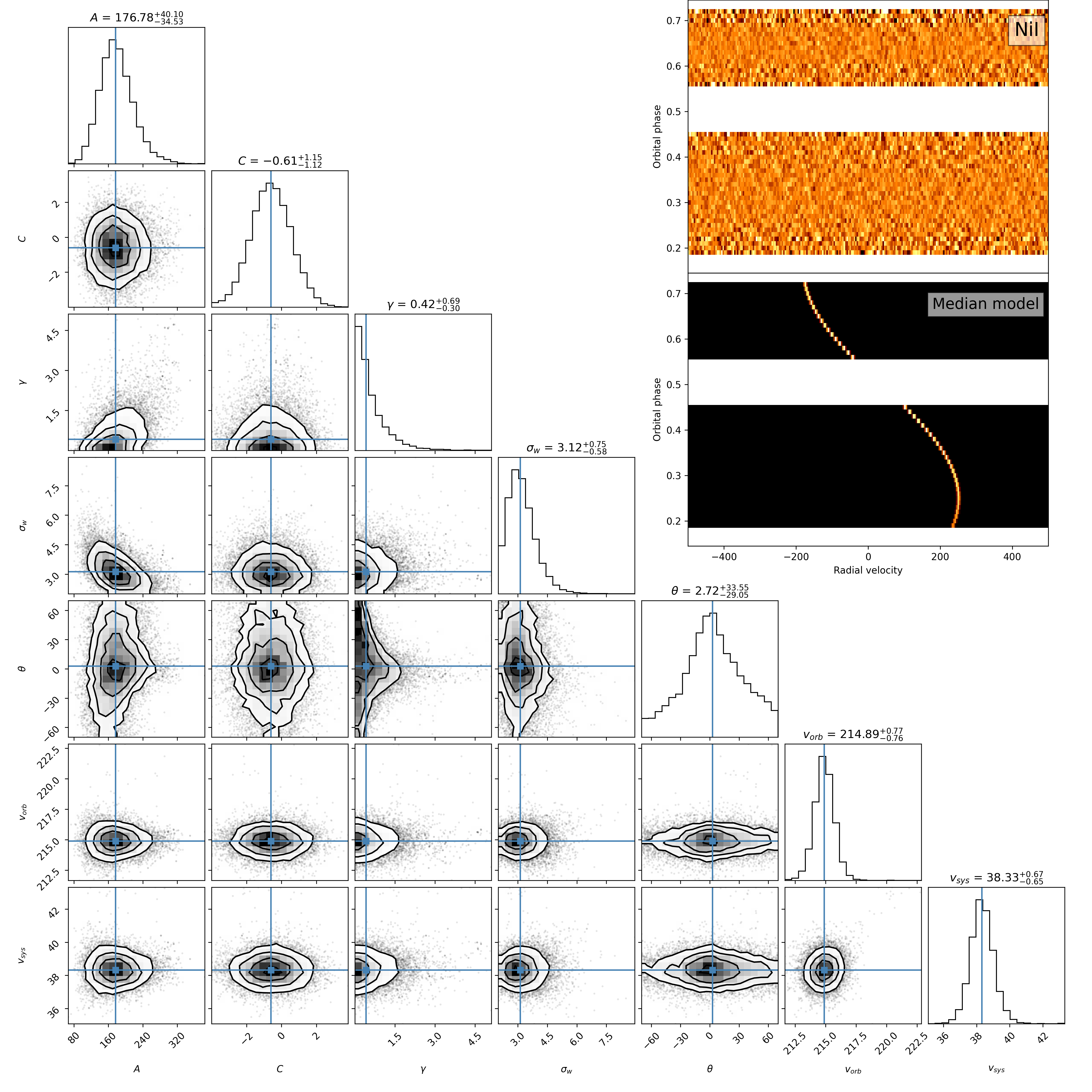}
\caption{Same as Fig. \ref{Fig:posteriors}, for \ion{Ni}{I}.}
\end{figure*}

\begin{figure*}
\includegraphics[width=\textwidth]{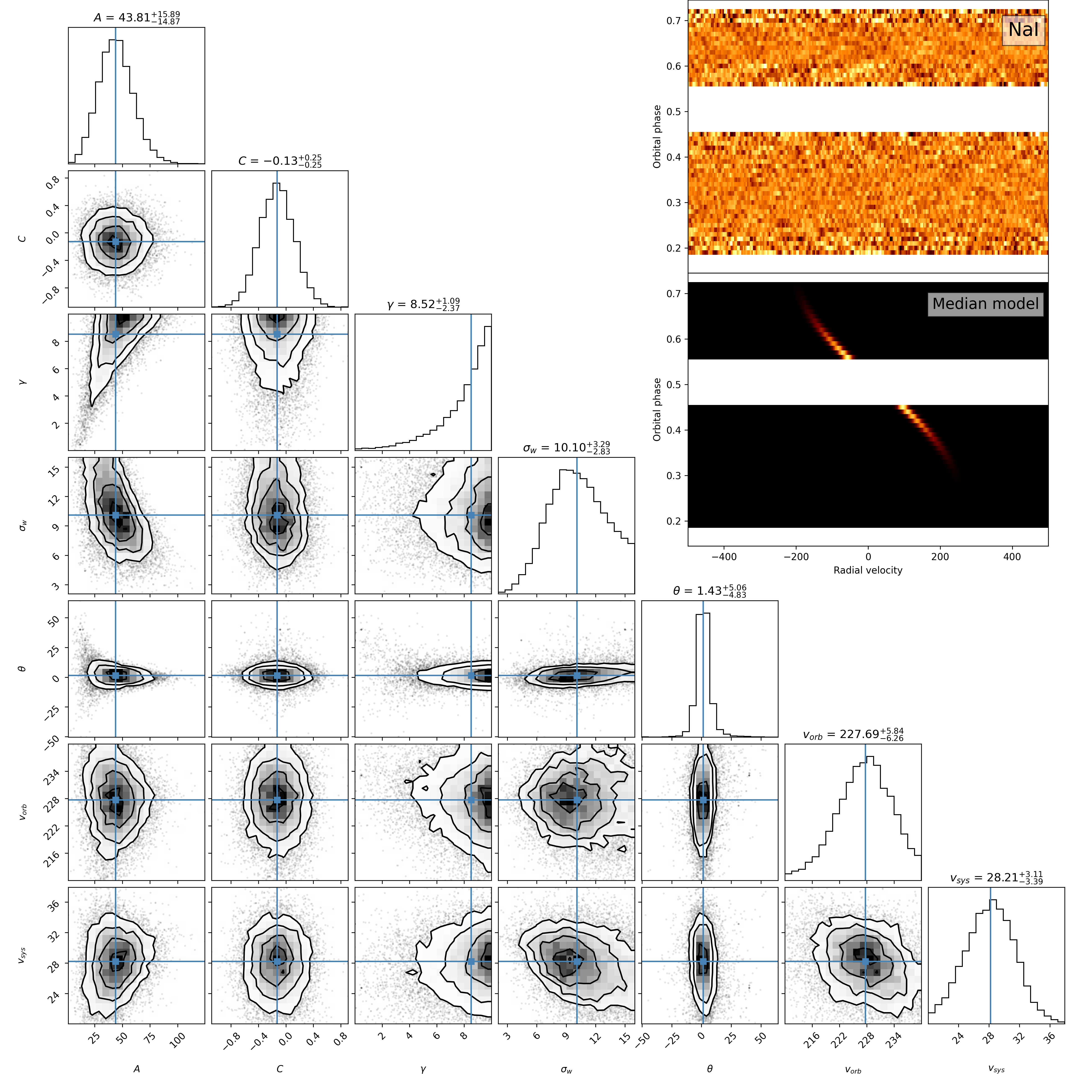}
\caption{Same as Fig. \ref{Fig:posteriors}, for \ion{Na}{I}.}
\end{figure*}

\end{appendix}
\end{document}